\documentclass[final,12pt,3p]{elsarticle}

\usepackage{graphicx,stfloats}
\usepackage{color}
\usepackage[colorlinks=true]{hyperref}
\usepackage{url}
\usepackage{blindtext}
\usepackage{latexsym,amsmath,amssymb}
\usepackage[T1]{fontenc}
\usepackage[utf8]{inputenc}
\usepackage[english]{babel}
\usepackage{csquotes}

\usepackage{algorithm2e}
\usepackage{scrextend}
\usepackage{mwe,tikz}
\usepackage[percent]{overpic}
\usepackage{color}
\usepackage[range-phrase={--},range-units=single]{siunitx}
\usepackage{graphicx}
\usepackage{tabularx}
\usepackage{subcaption}
\usepackage[export]{adjustbox}
\usepackage{wrapfig}
\usepackage{amsmath}
\graphicspath{{Figures/}}
\usepackage[version=4]{mhchem}
\usetikzlibrary{decorations.pathmorphing}
\tikzset{snake it/.style={decorate, decoration=snake}}

\usepackage[nameinlink, capitalize]{cleveref}
\newcommand{\ol}[1]{{\overline{#1}}}
\def\xvec{{\mbox{\boldmath$x$}}}

\usepackage{booktabs,multirow,multicol}
\usepackage{todonotes}
\biboptions{sort&compress}

\journal{Combustion and Flame}
\begin{document}
\begin{frontmatter}
\title{Data-assisted combustion simulations with dynamic submodel assignment using random forests}

\author[1]{Wai Tong Chung\corref{cor1}}
\ead{wtchung@stanford.edu}
\author[2]{Aashwin Ananda Mishra}
\author[1,3]{Nikolaos Perakis}
\author[1]{Matthias Ihme}

\address[1]{Department of Mechanical Engineering, Stanford University, Stanford, CA 94305, USA}
\address[2]{SLAC National Accelerator Laboratory, Menlo Park, CA 94025, USA}
\address[3]{Chair of Space Propulsion, Technical University of Munich, 85748 Garching,
Germany}

\cortext[cor1]{Corresponding author:}

\begin{abstract}
This investigation outlines a data-assisted approach that employs random forest classifiers for local and dynamic submodel assignment in turbulent-combustion simulations. This method is demonstrated in simulations of a single-element GOX/GCH4 rocket combustor; \emph{a priori} as well as \emph{a posteriori} assessments are conducted to (i) evaluate the accuracy and adjustability of the classifier for targeting different quantities of interest (QoIs), and (ii) assess improvements, resulting from the data-assisted combustion model assignment, in predicting target QoIs during simulation runtime. Results from the \emph{a priori} study show that random forests, trained with local flow properties as input variables and combustion model errors as training labels,
assign three different combustion models -- finite-rate chemistry (FRC), flamelet progress variable (FPV) model, and inert mixing (IM) -- with reasonable classification performance even when targeting multiple QoIs. Applications in \emph{a posteriori} studies demonstrate improved predictions from data-assisted simulations, in temperature and CO mass fraction, when compared with monolithic FPV calculations. 
An additional \emph{a posteriori} data-assisted simulation of a modified configuration demonstrates that the present approach can be successfully applied to different configurations, as long as thermophysical behavior can be represented by the training data.  
These results demonstrate that this data-driven framework holds promise for dynamic combustion submodel assignments in reacting flow simulations.
\end{abstract}


\end{frontmatter}
\section{Introduction}\label{Introduction}
High-fidelity simulations of turbulent reacting flows can incur high computational costs due to the  complexity required for employing  finite-rate chemical mechanisms and resolving relevant scales.
Numerous strategies \cite{LU2009192}  have been employed  for reducing the computational cost of detailed chemical mechanisms, such as (i) removing non-essential species and reactions \cite{Turnyi1990ReductionOL,Whitehouse2004}, (ii) lumping similar species and reaction pathways \cite{LI19891413,fournet2000}, (iii) time-scale analysis \cite{LU20011445,MAAS1992239}, and (iv) stiffness reduction \cite{Schwer2003,Singer2004}.

Alternatively, a  significant portion of combustion research has been devoted to the development of cost-efficient models for representing the combustion chemistry and turbulent scales~\cite{POPE_PCI2013}. 
The most popular of these low-order manifold models are categorized under flamelet methods, which represent combustion chemistry through solutions of representative flame configurations, such as laminar counterflow diffusion flames, freely propagating premixed flames, or homogeneous reactor systems. Examples of flamelet methods include the Burke-Schumann solution \cite{Burke1928}, the flame-prolongation in intrinsic lower-dimensional
manifold (FPI) \cite{GICQUEL20001901},
the flamelet-generated manifold (FGM) method \cite{Oijen2000}, and the flamelet/progress variable (FPV) method \cite{Pierce2004Progress-variableCombustion,IHME_CHA_PTISCH_PCI2005}.
These reduced manifold models are commonly employed to describe specific combustion regimes -- a multitude of which can exist within practical combustors. However, expert knowledge and experimental data is often required to correctly assign the most appropriate combustion model.

One solution to this issue is provided by dynamic adaptive chemistry methods~\cite{Liang2015AMethods,xie2017dynamic,Yang2017ParallelFlame} that save computational cost by reducing detailed chemical mechanisms, and transitioning between smaller sets of chemical models to represent combustion regimes of different chemical fidelity. A general mathematical framework was proposed by~\citet{WU_SEE_WANG_IHME_CF2015,Wu2019Pareto-efficientFlame} through the Pareto-efficient combustion (PEC) approach. In this approach, the compliance of a combustion submodel with the underlying flow-field representation is assessed through the construction of a so-called drift term, taking into consideration user-specific requirements about quantities of interest (QoI) and computational cost~\cite{DOUASBIN_IHME_ARNDT_CF2020}. While mathematically rigorous, these techniques are limited by their reliance on local information regarding the chemical composition and the construction of the model-compliance indicator. In contrast, data-driven methods can potentially offer a universal solution by allowing for the consideration of a wider range of conditions and processes that cannot be easily represented in the form of mathematical expressions accessible to an indicator function.

Data-driven methods involve the extraction of knowledge from data \cite{Dhar2013DataPrediction}.
These methods can be useful as long as a  substantial corpus of data is available to infer relationships between input variables and QoIs. 
As such, the employment of learning algorithms are gaining popularity in the simulation of turbulent flows. These methods have shown success in quantifying uncertainty \cite{Ling2015EvaluationUncertainty}, and augmenting closure models \cite{Wang2016Data-drivenStructures,Wu2018Physics-informedFramework} in Reynolds-averaged Navier–Stokes (RANS) simulations and large eddy simulations (LES) \cite{Duraisamy2019}.

In simulations of turbulent reacting flows, data-driven methods have also been applied with generating additional subgrid-scale closure that arises from the inclusion of combustion chemistry.  In particular, artificial neural networks have been employed for regressing thermophysical quantities in LES of turbulent flames~\cite{CHRISTO_MASRI_NEBOT_POPE_PCI1996,BLASCO_FUEYO_LARROYA_DOPAZO_CHEN_CCE1999,Ihme2009OptimalFlame,KEMPF_FLEMMING_JANICKA_PCI2005,SEN_MENON_CF2010}.  \emph{A priori} studies have been performed to demonstrate that convolutional neural-networks can provide accurate closure for turbulent combustion models~\cite{Lapeyre2019TrainingRates}. \citet{Ranade2019AValidation} conducted an \textit{a posteriori} study to show that artificial neural networks (ANNs) can be trained with experimental data to  generate closure models for chemical scalars in RANS simulations of turbulent jet flames.
\citet{DEFRAHAN_SHASHANK_KING_DAY_GROUT_CF2019} evaluated the use of ANNs, random forests, and generative learning methods for predicting the sub-filter probability density function in a turbulent combustion LES. \citet{Seltz2019DirectNetworks}  employed convolutional neural networks to generate closure for  unresolved terms in  the filtered progress variable transport equation. \citet{Yao2020} demonstrated that ANNs can be used to approximate the conditional scalar dissipation rate  in spray flame LES. 

To reduce computational costs that arise from complex combustion chemistry, various strategies have been employed through learning algorithms. 
Artificial neural networks were first successfully integrated within simulations of turbulent reacting flows as an alternative for representing chemical reactions  \cite{CHRISTO_MASRI_NEBOT_POPE_PCI1996,Blasco1998,BLASCO_FUEYO_LARROYA_DOPAZO_CHEN_CCE1999}.
\citet{CHATZOPOULOS20131465}, and \citet{Franke2017} demonstrated that training data extracted from 100 laminar flamelets was sufficient for training ANNs for representing chemistry in simulations more complex turbulent flame configurations. With this generic training set, ANNs showed a small capacity for extrapolation, but it was noted that accurate predictions were challenging if the target predictions deviated too largely away from the training set. 
\citet{SEN_MENON_CF2010}, and \citet{Alqahtani2021} also demonstrated that ANNs can be used for replacing stiff ODE solvers in turbulent flame simulations, with good accuracy and CPU performance. 
\citet{Ihme2009OptimalFlame,KEMPF_FLEMMING_JANICKA_PCI2005}, and \citet{OWOYELE_KUNDU_AMEEN_ECHEKKI_SOM_IJER_2020} used optimal ANN tabulation to replace conventional tabulation methods in manifold-based simulations.

These aforementioned approaches typically involve the use of regression for estimating numerical predictions. 
 Regression models in flow-physics problems are still in its infancy, and face challenges when extrapolating without an appropriate training set -- resulting in errors that arise from  numerical predictions that only match specific flow configurations represented by the training data  \cite{Wu2018Physics-informedFramework}.
  The present study ameliorates this issue by employing a classification algorithm that assigns well-tested physics-based combustion submodels of varying fidelity and complexity within the simulation domain. 
  Thus, the potential approximation errors made by the machine-learning algorithm are limited by the predictive capability of the lowest performing submodel.

 In the approach that is proposed in this work, local thermophysical quantities in the flow field are utilized as features for a random forest algorithm that spatially and dynamically assigns combustion submodels. 
 Random forests are an ensemble learning method commonly used in both classification and regression problems.
 Errors made by submodels, when predicting user-defined QoI, are used to construct the labels used for training the random forest.  
 Overall computational fidelity and cost of the simulation is determined by a user-defined submodel error threshold during training. This approach couples the assigned combustion submodels in the \emph{a posteriori} simulations by employing the mass-conserving approach developed by~\citet{Wu2019Pareto-efficientFlame}, but with a data-driven assignment approach that replaces the drift-term in the original PEC formulation.

This investigation is performed with the following objectives:\begin{itemize}
    \item To introduce classification algorithms for combustion submodel assignment, and assess the resulting data-assisted simulations.
    \item To evaluate  the suitability, accuracy, and adjustability of  random forests for submodel assignment. 
\end{itemize}
To this end, random forests are assessed for the purpose of local and dynamic model assignment in simulations of a gaseous-oxygen/gaseous-methane (GOX/GCH4) single-element rocket combustor~\cite{Silvestri2015ComparisonSections,Silvestri2016InvestigationChamber}. The mathematical models for simulating the turbulent combustion are presented in~\cref{sec:mathModel}. The experimental configuration, computational setup and baseline simulations using monolithic combustion models are discussed in~\cref{SEC_EXP_COMP_SETUP}. The  data-driven framework is  introduced in~\cref{sec:method}. Results from \emph{a priori} and \emph{a posteriori} assessments of the random forests are presented and discussed in~\cref{sec:results}, before offering concluding remarks in~\cref{sec:conclusions}.
\section{\label{sec:mathModel}Mathematical models}
\subsection{\label{sec:framework_numMethods}Computational method}
\label{sec:numeric}
The governing equations that are solved in the present study are the Favre-filtered conservation equations for mass, momentum, energy, and chemical species:
\begin{subequations}
\label{EQ_GOVERNING}
\begin{align}
    \partial_t \ol{\rho} + \nabla \cdot (\ol{\rho} \widetilde{\boldsymbol{u}} ) &= 0\\
    \partial_t (\ol{\rho} \widetilde{\boldsymbol{u}}) + \nabla \cdot (\ol{\rho} \widetilde{\boldsymbol{u}} \widetilde{\boldsymbol{u}} ) &= 
    - \nabla \cdot (\ol{p}\boldsymbol{I}) 
    + \nabla \cdot( \ol{\boldsymbol{\tau}}_{v}+  {\boldsymbol{\tau}}_{t})\\
    \partial_t (\ol{\rho} \widetilde{e}) + \nabla \cdot [\widetilde{\boldsymbol{u}} (\ol{\rho}  \widetilde{e} + \ol{p})] &= 
    - \nabla \cdot( \ol{\boldsymbol{q}}_{v}+ {\boldsymbol{q}}_{t})+\nabla \cdot [ 
    (\ol{\boldsymbol{\tau}}_{v} + {\boldsymbol{\tau}}_{t} )\cdot \widetilde{\boldsymbol{u}}] \\
    \partial_t (\ol{\rho} \widetilde{\boldsymbol{\phi}}) + \nabla \cdot (\ol{\rho} \widetilde{\boldsymbol{u}} \widetilde{\boldsymbol{\phi}}) &= 
    -\nabla \cdot( \boldsymbol{\ol{J}}_{ v} + \boldsymbol{{J}}_{ t})+ \ol{\dot{\boldsymbol{S}}} \label{eq:speciesTrans}
\end{align}
\end{subequations}
with density $\rho$, velocity vector $\boldsymbol{u}$,  specific total energy $e$, stress tensor $\boldsymbol{\tau}$, and  heat flux vector $\boldsymbol{q}$; $\ol{\,\cdot\,}$ denotes a filtered quantity and $\widetilde{\,\cdot\,}$ is a Favre-filtered quantity. Subscripts $v$ and $t$ denote viscous and turbulent quantities, respectively. Pressure $p$ is computed from the ideal gas equation of state.  $\boldsymbol{\phi}$, $\boldsymbol{J}$, and $\dot{\boldsymbol{S}}$ are the transported scalars, scalar diffusive flux, and scalar source term for the candidate combustion models.
Molecular fluxes are modeled using the mixture-averaged diffusion model.
The combustion models that are employed in the present study are described in detail in~\cref{sec: framework_combModel}.
 
Simulations are performed by employing an unstructured compressible finite-volume solver~\cite{Khalighi_2011,MA_LV_IHME_JCP2017,Wu2019Pareto-efficientFlame}. A central scheme, which is 4th-order accurate on uniform meshes, is used along with a 2nd-order ENO scheme. 
The ENO scheme is activated only in regions of high local density variation using a threshold-based sensor. 
A Strang-splitting scheme is employed for time-advancement, combining a strong stability preserving 3rd-order Runge-Kutta (SSP-RK3) scheme for integrating the non-stiff operators with a semi-implicit Rosenbrock-Krylov scheme \cite{Wu2019EfficientChemistry} for advancing the chemical source terms.  The dynamic Smagorinsky model~\cite{Moin1991ATransport} is used as closure for the subgrid-scale stresses. Turbulence/chemistry interaction is accounted for using the dynamic thickened-flame model~\cite{Colin2000}, employing a maximum thickening factor of 3, which is estimated through 1D flame calculations \emph{a priori}. Outside the flame region, both turbulent Prandtl and Schmidt numbers are prescribed at constant values of 0.7.
\subsection{\label{sec: framework_combModel}Combustion models} 
 In this work, we perform LES calculations that employ three different combustion submodels, namely a finite-rate chemistry (FRC) model, the flamelet/progress variable (FPV) model~\cite{Pierce2004Progress-variableCombustion,IHME_CHA_PTISCH_PCI2005}, and  an inert mixing (IM) model. The FRC model is defined by solving the species transport equation, \cref{eq:speciesTrans}, through direct integration. This method does not rely on strong assumptions on flame structure and is suitable for representing complex flows as well as intermediate species and unsteady effects. Despite the high-fidelity offered by FRC, since the cost of evaluating the chemical source terms scale linearly with the number of species, the utilization of  a large chemical mechanism can be prohibitively costly.  FPV approach aims to alleviate the computational cost of combustion chemistry by representing the thermochemical state space using a low-dimensional manifold based on flamelets, a series of one-dimensional diffusion flames. FPV relies on the observation that laminar diffusion flames are weakly affected by the presence of turbulence, which allows the turbulent diffusion flame to be represented by flamelets. While FPV is computationally efficient, it assumes adiabaticity and cannot model effects of heat-flux across boundaries well.  Lastly, IM models can only consider mixing without combustion chemistry.
 
The representation of scalar $\widetilde{\boldsymbol{\phi}}$  between FRC and the two tabulated chemistry models is dissimilar: FRC uses a chemical state-vector $\widetilde{\boldsymbol{\phi}} = [\widetilde{Y}_1, \ldots,\widetilde{Y}_{N_S}]^T$, consisting of $N_S$ number of chemical species, while the FPV and IM state-vector  is represented in terms of a low-dimensional manifold $\boldsymbol{\widetilde{\phi}}={\cal M}(\boldsymbol{\widetilde{\psi}})$, 
where $\boldsymbol{\widetilde{\psi}}$  is
the state vector that is used to parameterize the manifold. 
With the flame being artificially thickened as discussed in  \Cref{sec:numeric}, FPV is parameterized by the mixture fraction and progress variable $\widetilde{\boldsymbol{\psi}} = [\widetilde{Z},\widetilde{C}]^T$  which  differs from the conventional practice of using a presumed-PDF closure~\cite{Wu2019Pareto-efficientFlame}. The progress variable is defined as a linear combination of species mass fractions~\cite{IHME_SHUNN_ZHANG_JCP2012}: $C=Y_{\ce{CO2}}+Y_{\ce{H2O}}+Y_{\ce{CO}}+Y_{\ce{H2}}.$ For an inert and adiabatic mixture, the thermochemical state is fully parameterized by a single scalar, $\boldsymbol{\widetilde{\psi}} = [\widetilde{Z}]$. 

The present framework resolves the discrepancy in scalar representation when coupling different combustion models with the approach developed by~\citet{Wu2019Pareto-efficientFlame}. In this approach, a transport equation for mixture fraction is solved holistically in all models. Reconstruction of the chemical state-vector needed for FRC involves interpolation from the chemistry tables that stores all species, whereas the reconstruction of the progress variable needed for tabulated chemistry involves the sum of all major combustion product species: \ce{CO2}, \ce{CO}, \ce{H2O}, and \ce{H2}. To ensure consistency between the submodels, the aforementioned reconstruction is applied for the inactive combustion model at the submodel interface at every timestep.
Since the conservation laws for mass, momentum, and energy are universal among all combustion submodels, these properties are conserved throughout the domain. 
In addition, the choice of the dynamically-thickened flame model for the FRC and both manifold-based models avoids potential complications, since this closure model has been successfully applied to previous non-premixed flame simulations employing  FRC and tabulated chemistry models \cite{Wu2019Pareto-efficientFlame,FELDEN2018270,VREMAN2008394}.

The GRI-3.0 model~\cite{GRI30}, involving $N_S=33$ chemical species, is used to describe the reaction chemistry in all combustion models. FRC is incorporated into the LES solver using the Cantera library interface~\cite{Goodwin2018Cantera:Processes}. 
The molecular diffusion of chemical species is modeled with constant Lewis numbers, which are calculated at equilibrium condition of a stoichiometric \ce{CH4} and \ce{O2} mixture. 
The chemistry table employed in the FPV-model is constructed from the solution of steady-state counterflow diffusion flames that are solved in composition space~\cite{FLAMEMASTER}. 
 The Lewis numbers for the mixture fraction and progress variable are set at unity.

\section{\label{SEC_EXP_COMP_SETUP}Experimental configuration, computational setup and baseline simulations}
\subsection{\label{SSEC_EXP}Experimental configuration}
To evaluate the merit of the data-assisted classification method, we perform simulations of a single-element GOX/GCH4 rocket combustor~\cite{Silvestri2015ComparisonSections,Silvestri2016InvestigationChamber}. The  experimental configuration consists of a co-axial injector element where the oxidizer flows through a central jet with diameter $d_{o}=4\,$mm and the fuel is injected via an annulus with inner and outer diameters $d_{f,i}=5\,$mm and $d_{f,o}=6\,$mm. The combustion chamber with a total length of 285$\,$mm has a cylindrical shape with diameter $d_{ch}=12\,$mm. A conical nozzle is attached at the end of the combustion chamber, having a contraction ratio of 2.5. This setup results in a Mach number of approximately 0.25 in the combustion chamber, which is similar to typical flight configurations. The combustor operates at a nominal operating pressure of 20~bar and a global oxidizer-to-fuel ratio of 2.6, with mass flow rates of oxidizer $\dot{m}_o$ and fuel $\dot{m}_f$  measured at  34.82 g/s and  13.39 g/s, respectively. The temperature of the oxidizer and the fuel supplied at the injector inlet are $T_o =
 275\,\text{K}$ and $T_f = 269\,\text{K}$. Static wall pressure and wall heat flux are measured through thermocouples and pressure transducers, installed along the chamber wall. 
\subsection{\label{SSEC_COMP_SETUP}Computational setup}
\begin{figure}[!htb!]
 \centering
 \includegraphics[width=\textwidth]{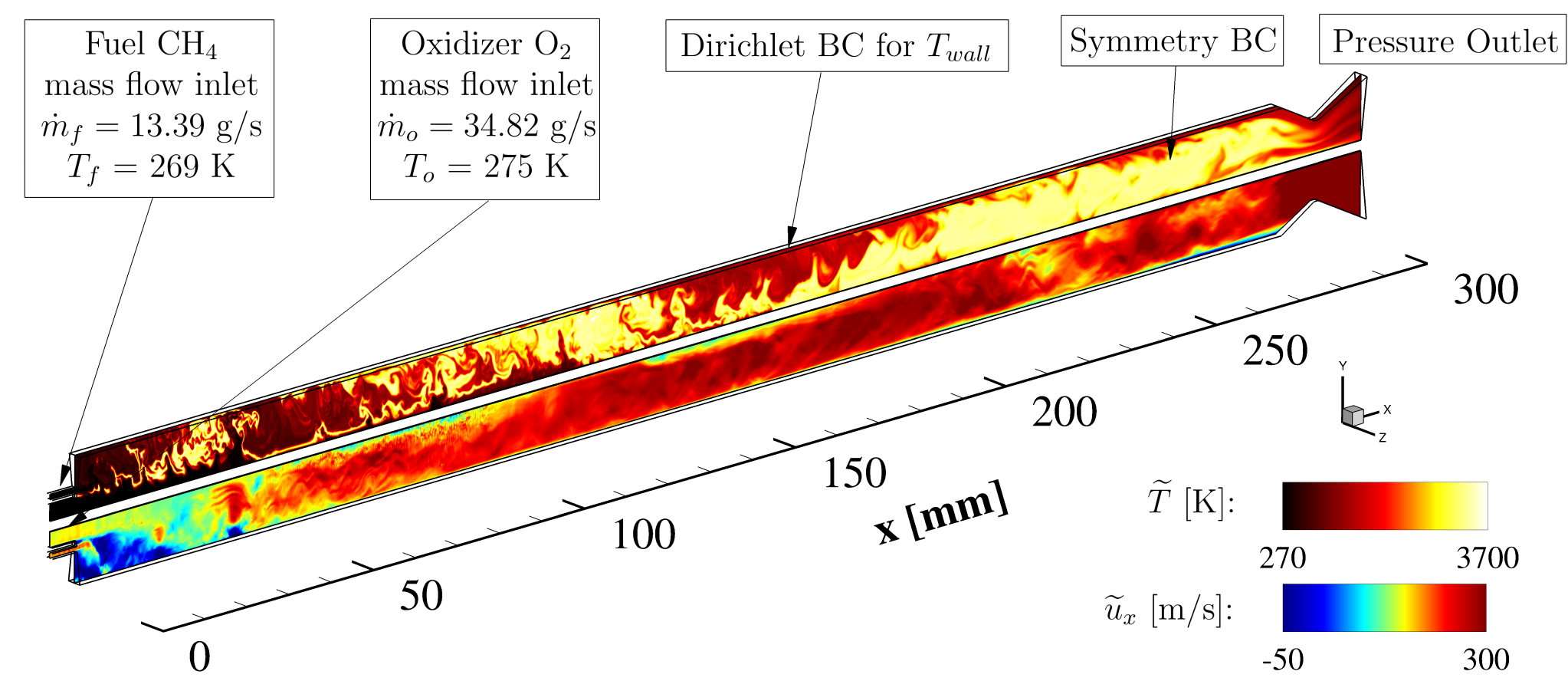}
 \caption{Computational domain presented in conjunction with instantaneous temperature (top) and axial velocity (bottom) fields from monolithic FRC simulations.}\label{fig:3d}
\end{figure}

In this model-assignment problem, we consider an axisymmetrical domain that is  representative of the single-element GOX/GCH4 rocket combustor, as shown in~\cref{fig:3d}. The domain consists of a $3^\circ$ combustor sector, with a truncation at 0.4 mm to remove the singularity at the centerline. Axisymmetric simulations of rocket combustors have been frequently employed to obtain insight in the turbulent combustion process~\cite{Zips2017Non-AdiabaticCombustion,Lapenna2018SimulationMethod}, while offering feasible computational costs. This was found to be crucial for the exploration of a wider range of parameters in the data-assisted method, especially with the use of a detailed FRC-model consisting of 33 chemical species in the present study.

At the inlets, the fuel and oxidizer mass flow rates and temperature are prescribed following the experimental measurements \cite{Silvestri2015ComparisonSections,Silvestri2016InvestigationChamber}. At the chamber and nozzle walls, the temperature profile is defined as a Dirichlet boundary condition, which is obtained from the measurements by~\citet{Perakis2019InverseChambers}. The bottom and axisymmetric faces are prescribed with symmetry boundary conditions.  All remaining boundaries are defined as adiabatic non-slip walls with the exception of the exhaust, which is modeled as a pressure outlet.  The computational domain is discretized by a  block-structured mesh consisting of $2\times10^5$ cells. The wall-normal direction is resolved down to 30 $\mu$m, and a wall model \cite{Kawai2013DynamicNumbers} is employed for the viscous sublayer.
Simulations are performed using 600 Intel Xeon (E5-2680v2) processors. The solution is advanced using a typical timestep of 25 ns, corresponding to a convective CFL number of 1.0.  
\subsection{\label{SSEC_MONO_LES}Baseline results from monolithic LES combustion simulations}
\begin{figure}[!htb!]
 \centering
 \begin{subfigure}[t]{\columnwidth}
  \centering
  \includegraphics[width = \columnwidth]{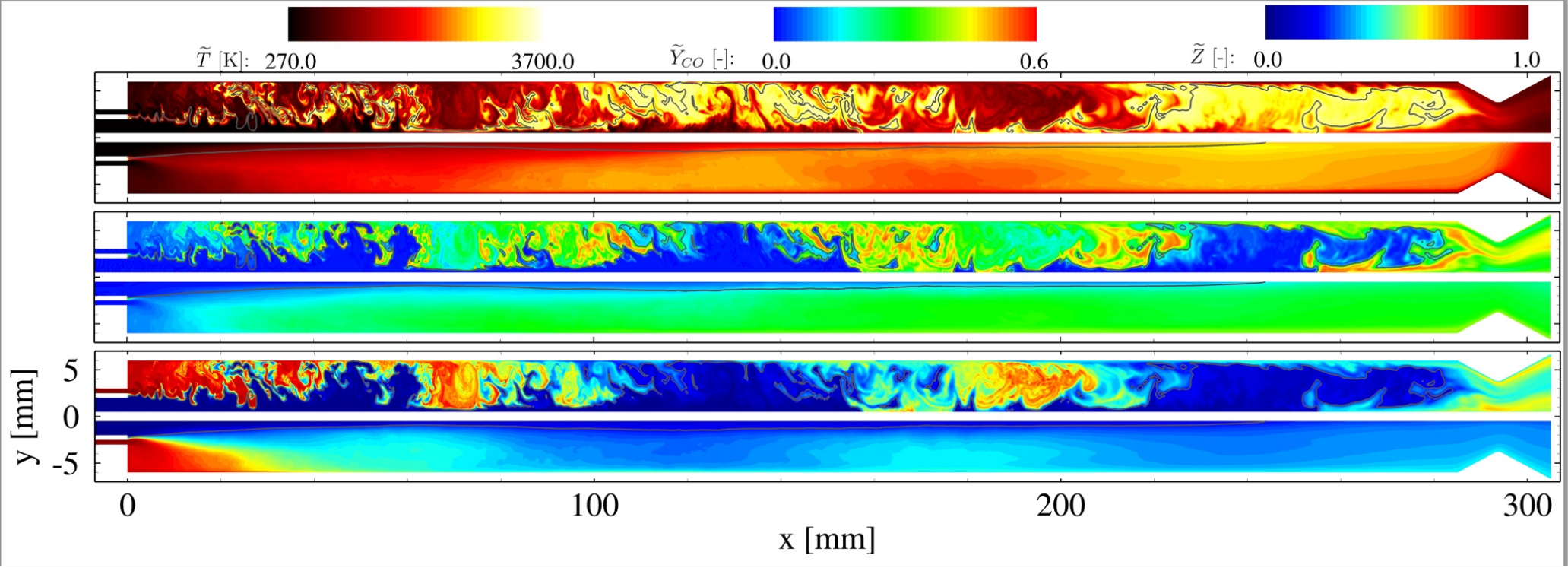}
  \caption{\label{fig:posterioriFRC}Monolithic FRC simulation.}
 \end{subfigure}
 \quad
 \begin{subfigure}[t]{1\columnwidth}
  \centering
  \includegraphics[width = \columnwidth]{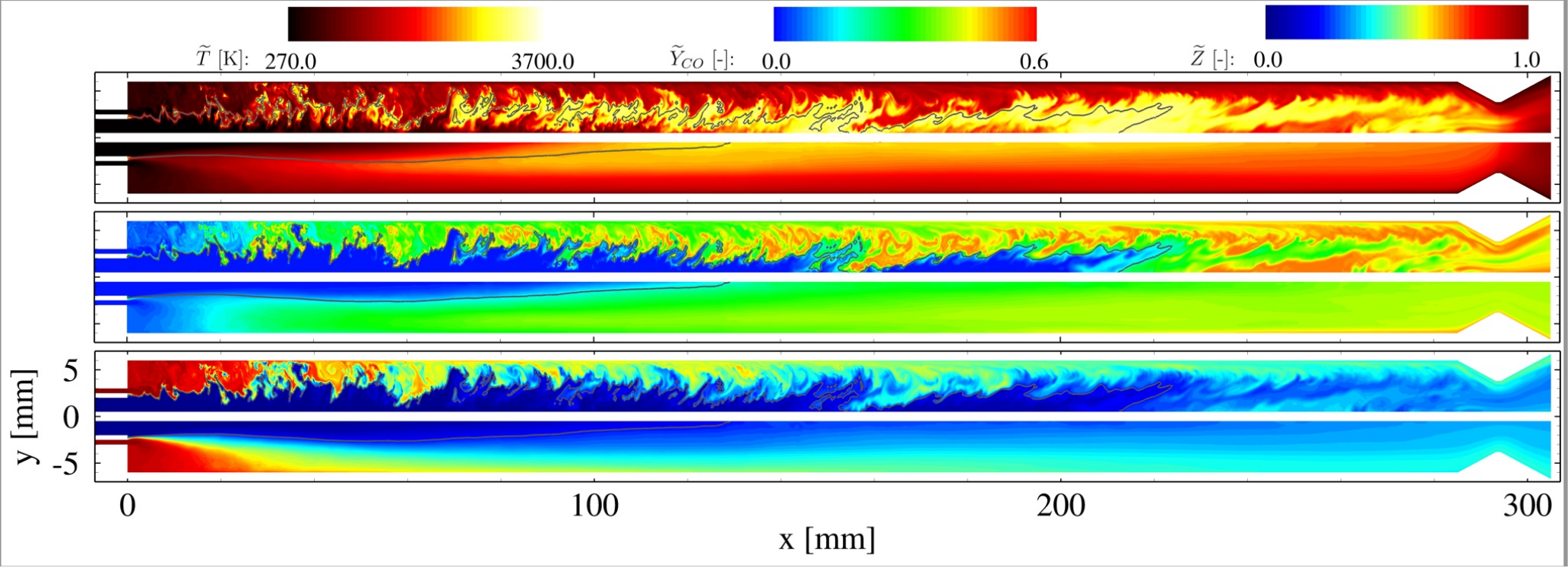}
  \caption{\label{fig:posterioriFPV}Monolithic FPV simulation.}
 \end{subfigure}
 \caption{\label{fig:posterioriFRC_FPV}Temperature, CO mass fraction, and mixture fraction fields (from top to bottom) for (a) monolithic FRC and (b) monolithic FPV simulations. Upper half: instantaneous fields, bottom half: time-averaged fields. The location of the stoichiometric mixture, $\widetilde{Z}_{st}=0.2$, is shown by black lines.}
\end{figure}

Simulations of the rocket combustor are first  performed using monolithic FRC and monolithic FPV simulations. Flow fields are initialized with equilibrium products and temperature, thus allowing the monolithic FPV simulation to ignite. Instantaneous and time-averaged fields of temperature, CO mass fraction, and mixture fraction from monolithic FRC calculations and monolithic FPV simulations are shown in~\cref{fig:posterioriFRC,fig:posterioriFPV}, respectively.  Results from the FRC simulations are qualitatively similar to previous simulations~\cite{Zips2017Non-AdiabaticCombustion,PERAKIS_HAIDN_IHME_PCI2021}, where a non-uniform mixture fraction field, a long oxygen core, and an agglomeration of cold rich gases to the chamber wall are observed. In contrast, some notable differences are observable from the FPV simulations, shown in~\cref{fig:posterioriFPV}. In particular, a thicker thermal boundary layer is seen for the FPV simulation. This difference is consistent with other LES studies~\cite{Ma2018NonadiabaticEngines} which have shown that an adiabatic FPV model, as employed in the present study, mispredicts the wall-heat loss and exothermic CO-recombination in the boundary layer~\cite{PERAKIS_HAIDN_IHME_PCI2021}. 

\section{\label{sec:method}Data-assisted simulation framework}
In this investigation, the present data-driven framework uses a supervised learning algorithm for combustion submodel assignment.  During training, the supervised learning algorithm learns a function $f: \xvec \mapsto y$ that maps  with data containing input vector $\xvec \in \mathcal{X}$, and the corresponding  true response $y \in \mathcal{Y}$. A trained supervised learning model can then provide an approximation for any output $y \in \mathcal{Y}$, when fed with a new input set $\xvec$.
The procedure for incorporating a supervised learning algorithm for combustion submodel assignment is as follows:

\begin{enumerate}
\item  Generate data either from experimental measurements or  numerical simulations. In this work, we use the instantaneous flow-field solutions from the FRC simulation of the GOX/CH4 rocket combustor as  learning dataset, discussed in~\cref{SEC_EXP_COMP_SETUP}.

\item Assign labels to the training data. Prior to training, each training datapoint is typically assigned a true response.     In this work, we present a multiclass classification problem for optimal assignment of three combustion models $ \mathcal{Y} = \{\text{IM, FPV, FRC}\}$. Hence, we use the local combustion submodel error of two essential local QoIs, namely $T$ and $Y_{\ce{CO}}$, to programmatically assign labels. Details are presented in~\cref{sec:framework_featurelabel}.
    
\item  Construct the feature vector $\xvec \in \mathcal{X}$. In this work, we apply a feature selection method based on the Maximal Information Coefficient (MIC) \cite{reshef2011detecting}, as discussed in~\cref{sec:feature}, to construct a feature set consisting of local thermophysical quantities that include the mixture fraction, progress variable, density, local Prandtl number, and Euclidean norm of the mixture fraction gradient, \emph{viz.,} $\xvec=[\widetilde{Z},\widetilde{C},\ol{\rho},\widetilde{T},Pr_{\Delta},\|\nabla \widetilde{Z}\|_2]$. 

\item Train, validate, and test the classification algorithm. In this work, a random forest classifier is used for combustion submodel assignment. Details of the algorithm are presented in~\cref{sec:framework_rf}.
\end{enumerate}

\subsection{Label assignment}
\label{sec:framework_featurelabel}
We present a multiclass classification problem for optimal assignment of three combustion models $ \mathcal{Y} = \{\text{IM, FPV, FRC}\}$. In this problem, we consider the FRC model as combustion model of highest fidelity but at the expense of highest computational cost. Hence, regions with local scalar predictions by IM and FPV models that match those of FRC can be considered optimally assigned. Therefore, we assign labels in the training set based on the normalized combustion submodel error  $\epsilon^{y}_{Q}$ of quantities of interest $\alpha \in Q$ between FRC and the models of lower fidelity~\cite{WU_SEE_WANG_IHME_CF2015}:
 \begin{equation}
    \epsilon_{Q}^{y} = \sum_{\alpha \in Q} w_\alpha \frac{ |\alpha^{\text{FRC}} - \alpha^{y}|}{\|\alpha^{\text{FRC}}\|_{\infty}} \enskip \text{with} \enskip y \in \{\text{FPV},\text{IM}\}\;,
\end{equation}
where the error for considering $N$   quantities-of interest is a weighted linear combination of each individual submodel error. The weights for each QoI $w_\alpha$ is subject to the following constraints: $\sum_{\alpha \in Q}^N w_\alpha  = 1$ and $w_\alpha \geq 0$. In this study, the use of temperature and mass fractions of CO and OH as QoIs. In the combined use of both temperature and CO mass fraction, $Q = \{\widetilde{T},\widetilde{Y}_{\text{CO}}\}$, both QoIs are equally weighted: $w_{T} = 0.5$ and $w_{\text{CO}} = 0.5$. Similarly for the combined use of three QoIs $Q = \{\widetilde{T},\widetilde{Y}_{\text{CO}},\widetilde{Y}_{\text{OH}}\}$, all QoIs are equally weighted: $w_{T} = 0.33$, $w_{\text{CO}} = 0.33$, and  $w_{\text{OH}} = 0.33$.
Temperature $\widetilde{T}$ is chosen as a proxy to describe the combustion efficiency and engine performance. The CO mass fraction $\widetilde{Y}_{\ce{CO}}$ is chosen to challenge the  deficiencies of tabulation methods in capturing intermediate species~\cite{Wu2019Pareto-efficientFlame}. OH mass fraction  $\widetilde{Y}_{\ce{OH}}$ is selected since radical formation is essential in combustion phenomena.

 FRC data is used to reconstruct  FPV and IM  quantities of interest $\alpha \in Q$ by interpolating the generated flamelet tables using reconstructed values of mixture fraction and progress variable: 
\begin{equation}
\label{eq:apriori}
    \alpha^y \approx \alpha^y_{\text{table}}(\widetilde{Z}_{\text{FRC}},\widetilde{C}_{\text{FRC}})
    \enskip \text{where} \enskip y \in \{\text{FPV},\text{IM}\}\;.
\end{equation}
The mixture fraction is computed using Bilger's definition~\cite{Bilger1976TurbulentFlames}, while the progress variable is computed using the sum of major combustion products, as described in~\cref{sec: framework_combModel}. We must note that since  $\alpha^y$ is reconstructed from FRC data, the resulting error metric $\epsilon_Q^y$ is an approximation of the true errors between FRC and tabulated chemistry. However, the use of this error metric is well-justified since Bilger's mixture fraction and the sum of major combustion products are robust quantities for bridging FRC and tabulated methods.
Labels are assigned programmatically as demonstrated in~\cref{alg:label}. In this algorithm,  a model of higher fidelity is assigned when the QoI submodel error $\epsilon_{Q}^{y} $ exceeds a user-defined threshold $\theta_{Q}^{y} $, with FRC chosen when all conditions for selecting FPV and IM are not met. While $\theta_Q^{\text{FPV}}$ and  $\theta_Q^{\text{IM}}$ can be assigned distinct values, throughout this study we will explore cases that use the same threshold for both IM and FPV, \emph{viz.,} $\theta_Q^{\text{IM}}=\theta_Q^{\text{FPV}} = \theta_Q $ for simplicity.
\begin{algorithm}[!htb!]
  \uIf{$\epsilon_Q^{\text{IM}}< \theta_Q^{\text{IM}}$}{use inert mixing (IM)}
  \uElseIf{$\epsilon_Q^{\text{FPV}}< \theta_Q^{\text{FPV}}$}{use tabulated chemistry (FPV)}
  \Else{use finite-rate chemistry (FRC)}
 \caption{Assigning labels in the training set.\label{alg:label}}
\end{algorithm}

\subsection{Feature selection}  \label{sec:feature}

Adding uninformative features to the learning dataset can reduce accuracy and  computational efficiency of learning algorithms \cite{Li2017FeaturePerspective}.
Carrying out appropriate feature selection beforehand can improve the interpretability of the predictions of the trained model. To this end, feature selection can be used for identifying the most descriptive and discriminative features from the raw dataset to use as inputs for our learning algorithms. In this work, we select features from  local quantities and group parameters that can characterize the reacting flow, combustion state, and turbulence. 

For feature selection, we rely on the Maximal Information-based Non-parametric Exploration (MINE) tools~\cite{reshef2011detecting} that utilize mutual information between variable pairs to ascertain the strength of  relationships between variables based on instantaneous flow-field representations from  a monolithic FRC simulation. MINE utilizes the Maximum Information Coefficient (MIC) to ensure (i) generality, where the association between the variables are not limited to a particular form such as linear associations, and (ii) equitability, where the effect of noise on different relationships is similar.

\begin{figure}[htb!]
        \centering
            \begin{subfigure}{0.32\columnwidth}
        \centering
            \includegraphics[width=\columnwidth]{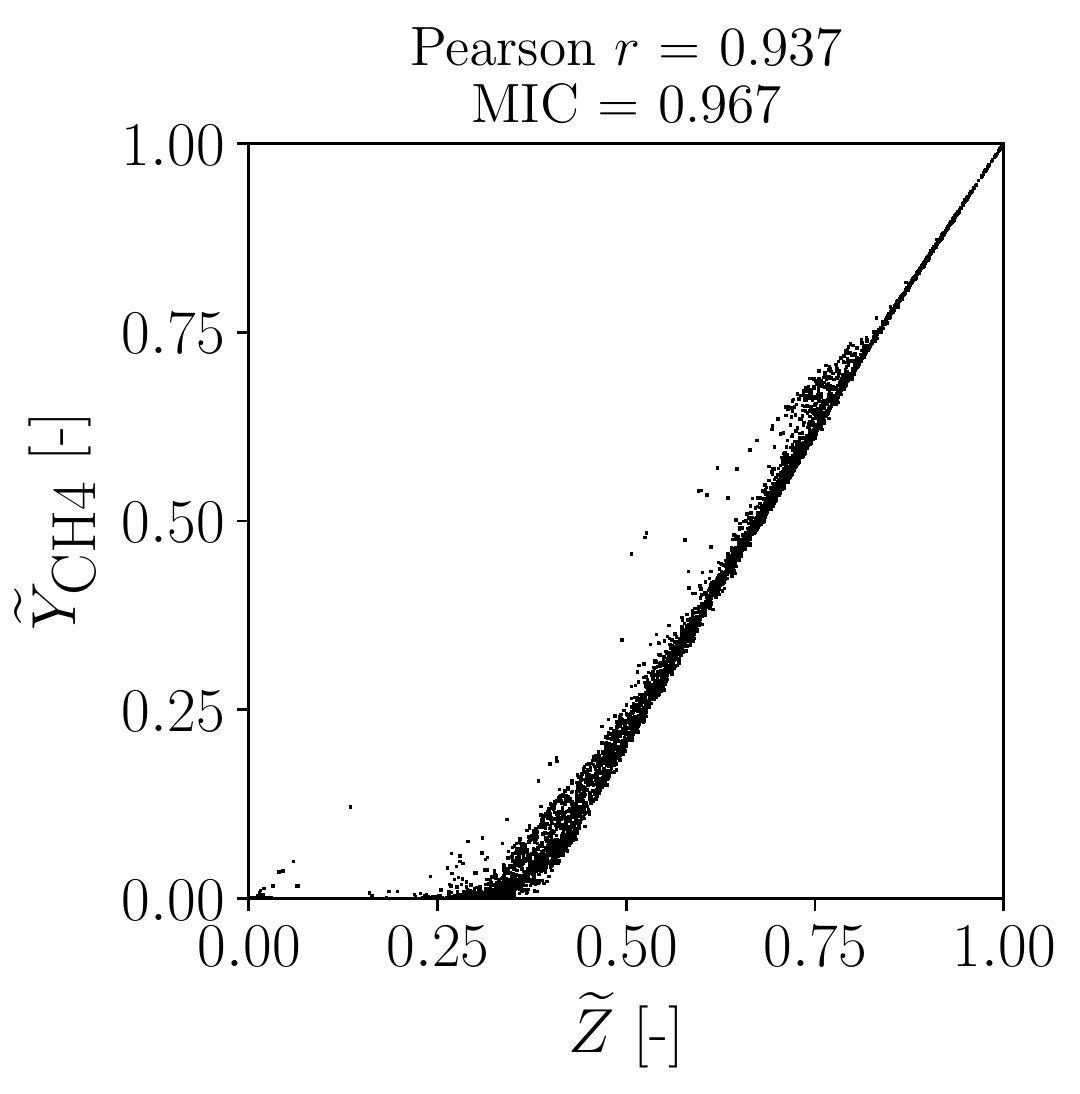}
        \caption{  \label{fig:ZCH4mic} }
    \end{subfigure}
    \begin{subfigure}{0.32\columnwidth}
        \centering
            \includegraphics[width=\columnwidth]{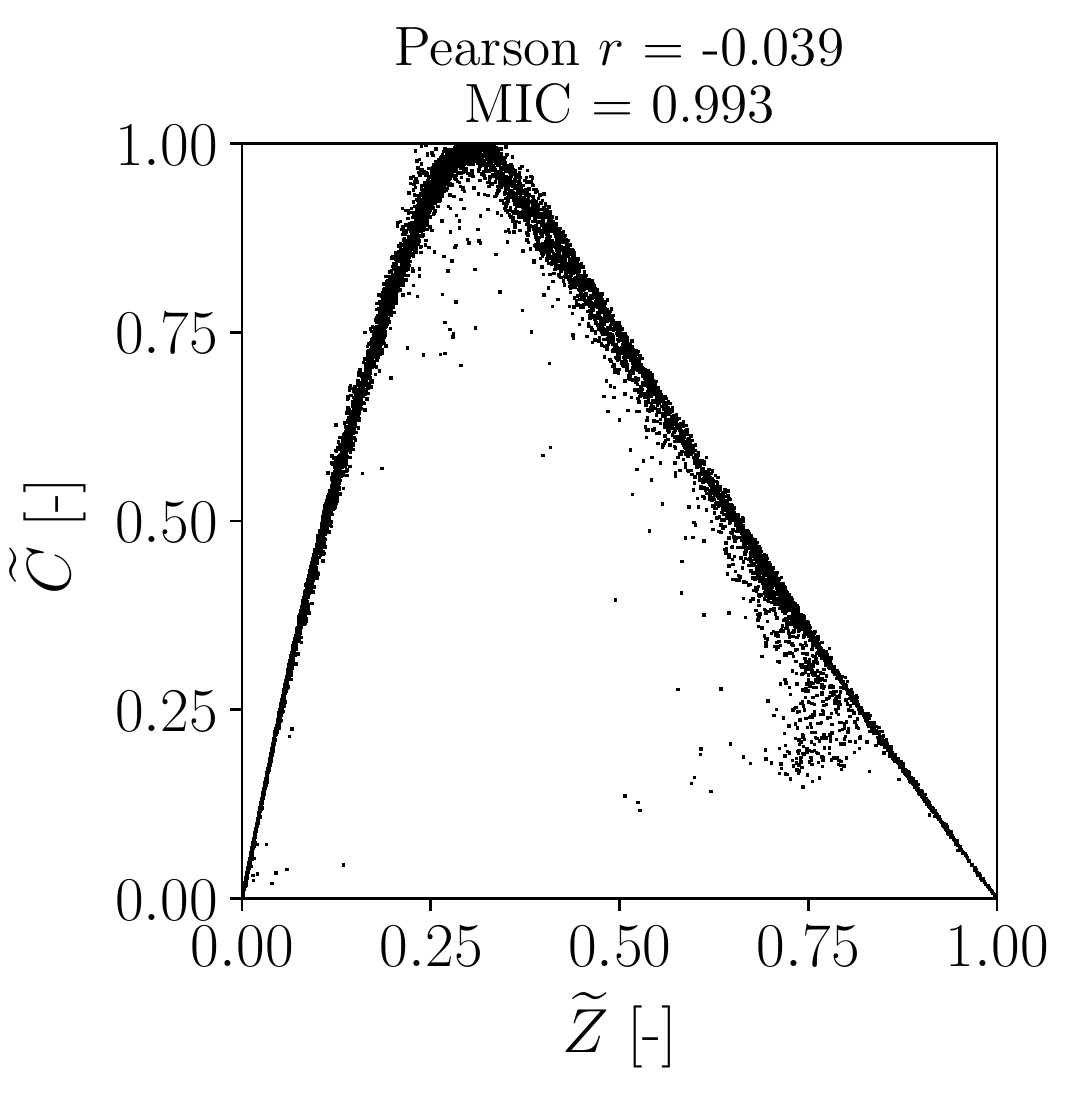}
            \caption{  \label{fig:ZCmic} }
    \end{subfigure}
    \begin{subfigure}{0.34\columnwidth}
        \centering
            \includegraphics[width=\columnwidth]{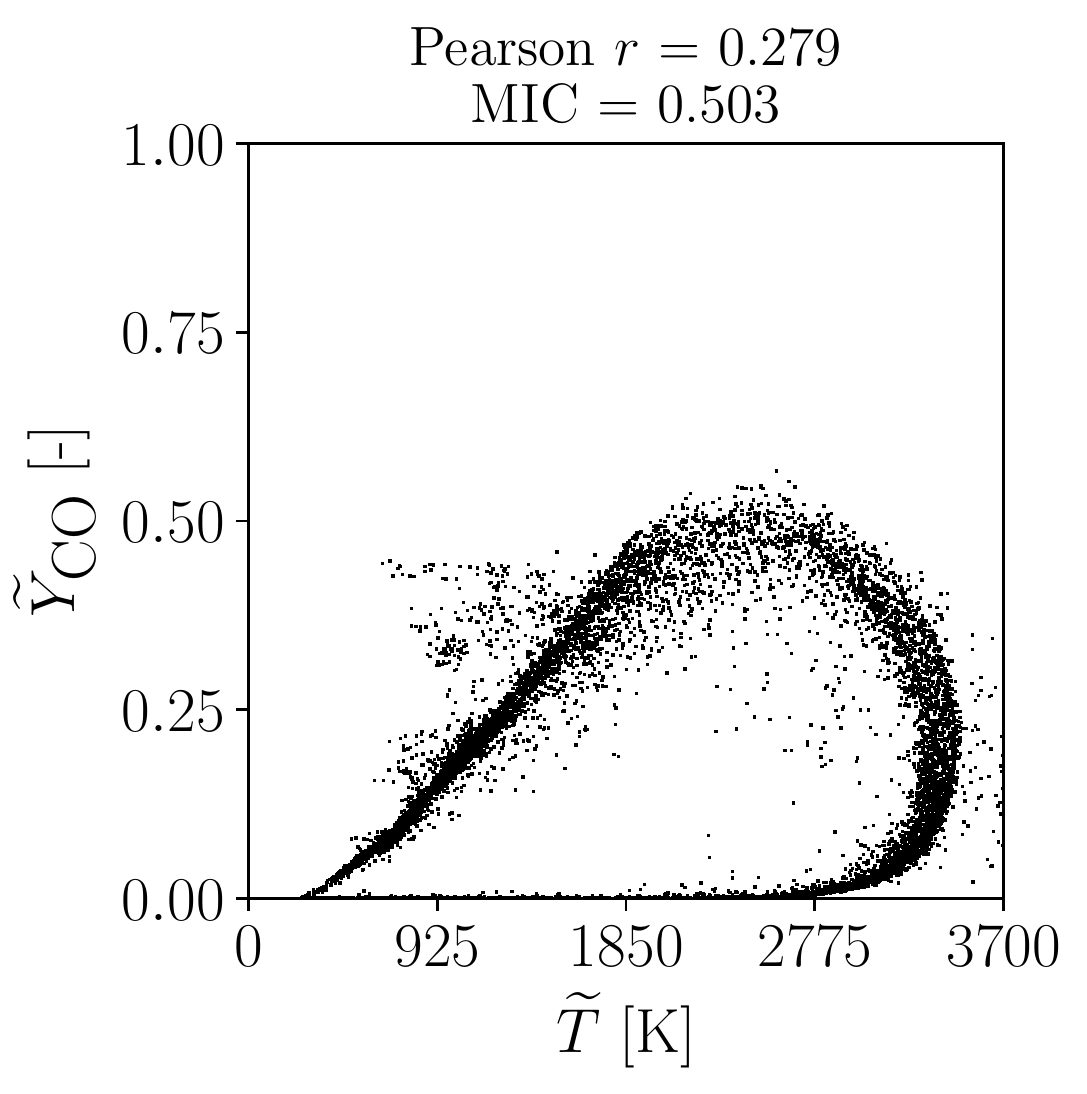}
            \caption{\label{fig:ZTmic} }
    \end{subfigure}
    
    \caption{ Comparison between Maximum Information Coefficient (MIC) and Pearson's Correlation Coefficient (Pearson $r$) for (a) near-linear scatter points, and (b,c) non-linear scatter points. \label{fig:MICComparison} }
\end{figure}

While Pearson's correlation has been utilized to ascertain the strength of relationships between variables in scientific applications, this does not account for any non-linear relationships. This is illustrated in~\cref{fig:MICComparison}, where Pearson's coefficient, or  Pearson $r$, is compared to MIC for different scatter points. As can be seen in \cref{fig:ZCH4mic}, for linear relationships with noise, both coefficients are similar. However, in \cref{fig:ZCmic,fig:ZTmic}, non-linear associations between variables are ignored by Pearson's correlation coefficient while MIC is able to account for such complex relationships. Mutual-information-based measures that ensure generality and equitability, like MIC, can be used to compare different features, rank them and select subsets of the most descriptive and discriminative features. Additionally, such mutual information based feature selection is model agnostic and can be used across different machine learning models, as a pre-processing step. In this vein,  MIC measure has been utilized for feature selection in prior works with success \cite{ge2016mctwo}.

\begin{figure}[!htb!]
  \centering
  \begin{subfigure}{\columnwidth}
    \centering
    \includegraphics[width=0.85\columnwidth]{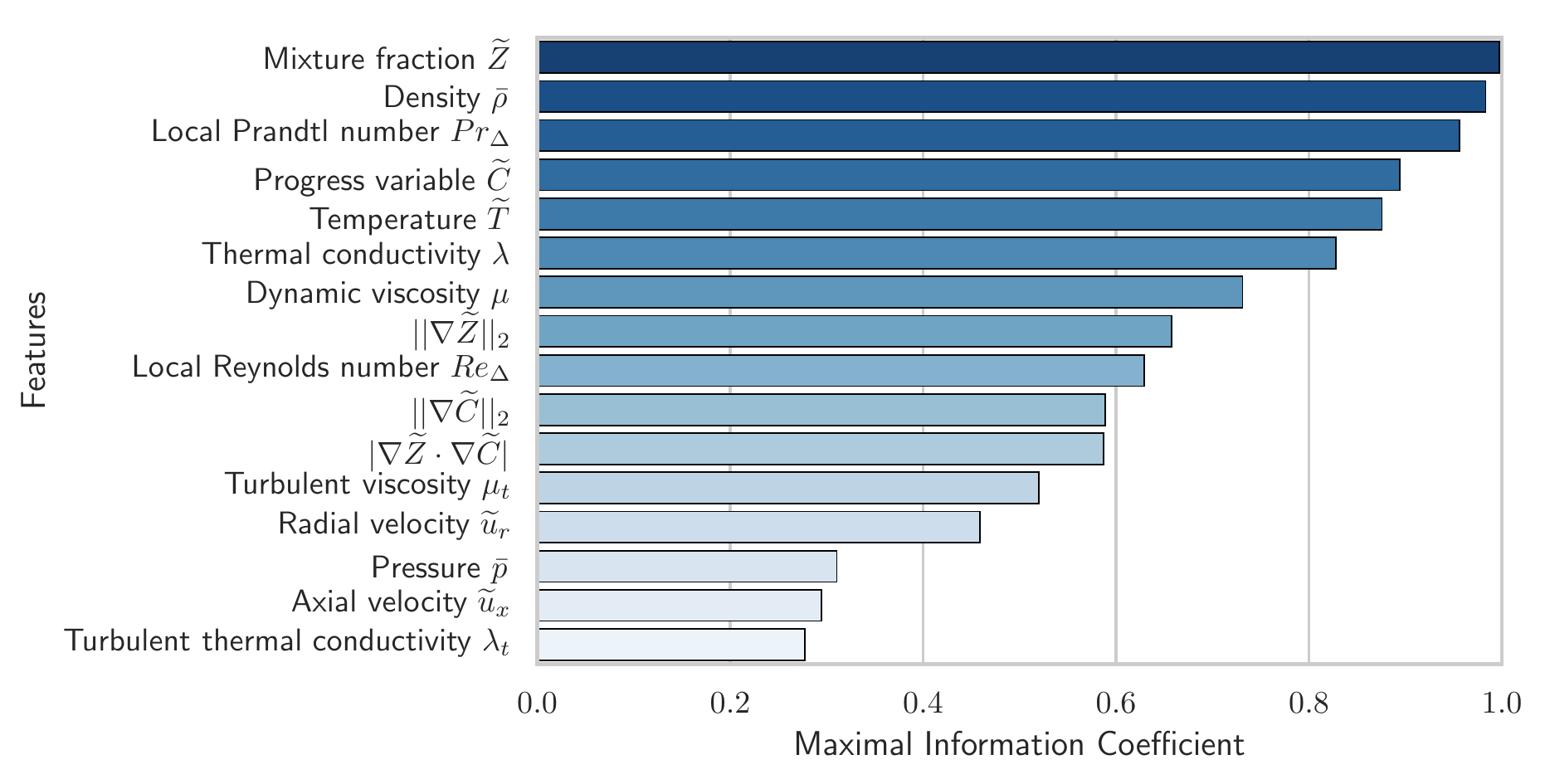}
    \caption{\label{fig:mic_IM}MIC relating features with IM model error $\epsilon_{\{T,\ce{CO}\}}^{\text{IM}}$.}
  \end{subfigure}
  \begin{subfigure}{\columnwidth}
    \centering
    \includegraphics[width=0.85\columnwidth]{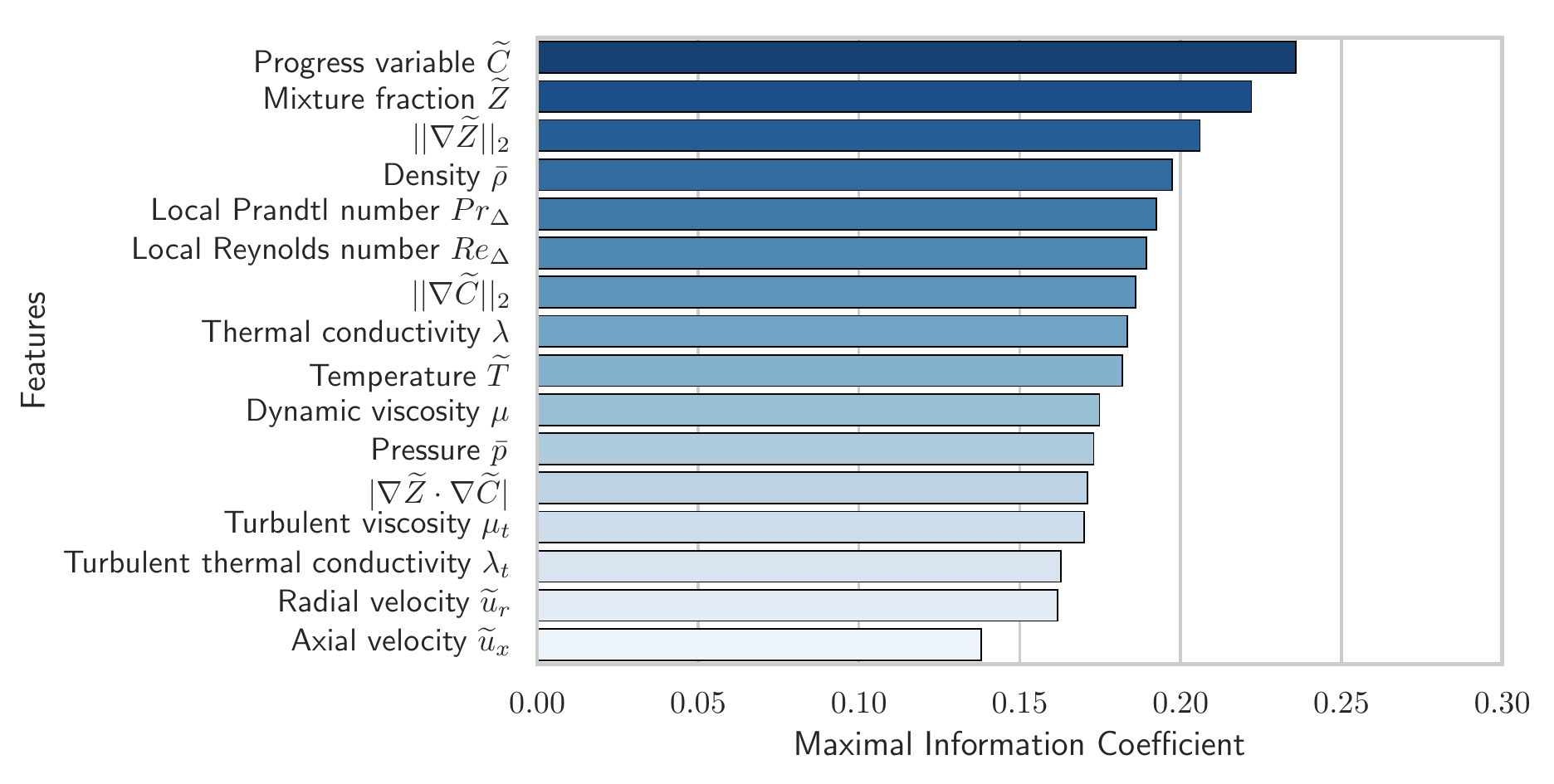}
    \caption{\label{fig:mic_FPV}MIC relating features with FPV model error $\epsilon_{\{T,\ce{CO}\}}^{\text{FPV}}$.}
  \end{subfigure}
  \caption{\label{fig:mic_IM_FPV}Maximal information coefficient score for features and model error.}
\end{figure}

\Cref{fig:mic_IM,fig:mic_FPV} show MIC scores relating 16 potential features with IM model error $\epsilon^\text{IM}_{\{T,\text{CO}\}}$ and FPV model error $\epsilon^\text{FPV}_{\{T,\text{CO}\}}$, respectively. These 16 potential features consist of thermophysical quantities and dimensionless quantities that characterize each cell within the domain. Dimensionless quantities include the local Prandtl number, $Pr_\Delta = \widetilde{\nu}/\widetilde{\alpha}$, comparing the local ratio of viscosity and thermal diffusivity, and the local Reynolds number, $Re_\Delta=\Delta |\widetilde{\boldsymbol{u}}|/\nu$, which is the ratio of inertial forces and viscous force within each cell and $\Delta$ denotes the characteristic length of each computational cell. It can be seen that the MIC scores for $\epsilon^\text{FPV}_{\{T,\text{CO}\}}$ are much lower than for $\epsilon^\text{IM}_{\{T,\text{CO}\}}$. This indicates that it is more challenging to form statistical relationships between features and FPV model errors than for IM model error. This observation is consistent with the intuition that it is much easier to identify failure of the IM models than the shortfall of the FPV model. 

In the following, the top five features from both MIC tests are used to construct the feature set consisting of mixture fraction, progress variable, density, local Prandtl number, and Euclidean norm of the mixture fraction gradient: $\xvec=[\widetilde{Z},\widetilde{C},\ol{\rho},\widetilde{T},Pr_{\Delta},\|\nabla \widetilde{Z}\|_2]^T$. The inclusion of $Pr_{\Delta}$ in the feature set is unexpected since $Pr_{\Delta}$ is approximately constant and has weak temperature dependence.
However, given that $Pr_{\Delta}$ is slightly higher in fuel and oxidizer when compared to combustion products, small variations within flow field prove useful for the random forests. 
We note that the data-driven framework in this study presently restricts the construction of feature and label sets to local quantities for simplicity.
More elaborate methods for incorporating spatial and temporal dependencies into present approach, through the use of convolutional neural networks \cite{Lapeyre2019TrainingRates,Seltz2019DirectNetworks}, should be subject to further study.
\subsection{Random forest classifier}
\label{sec:framework_rf}
\begin{figure}[!htb!]
 \centering
 \includegraphics[width=\textwidth]{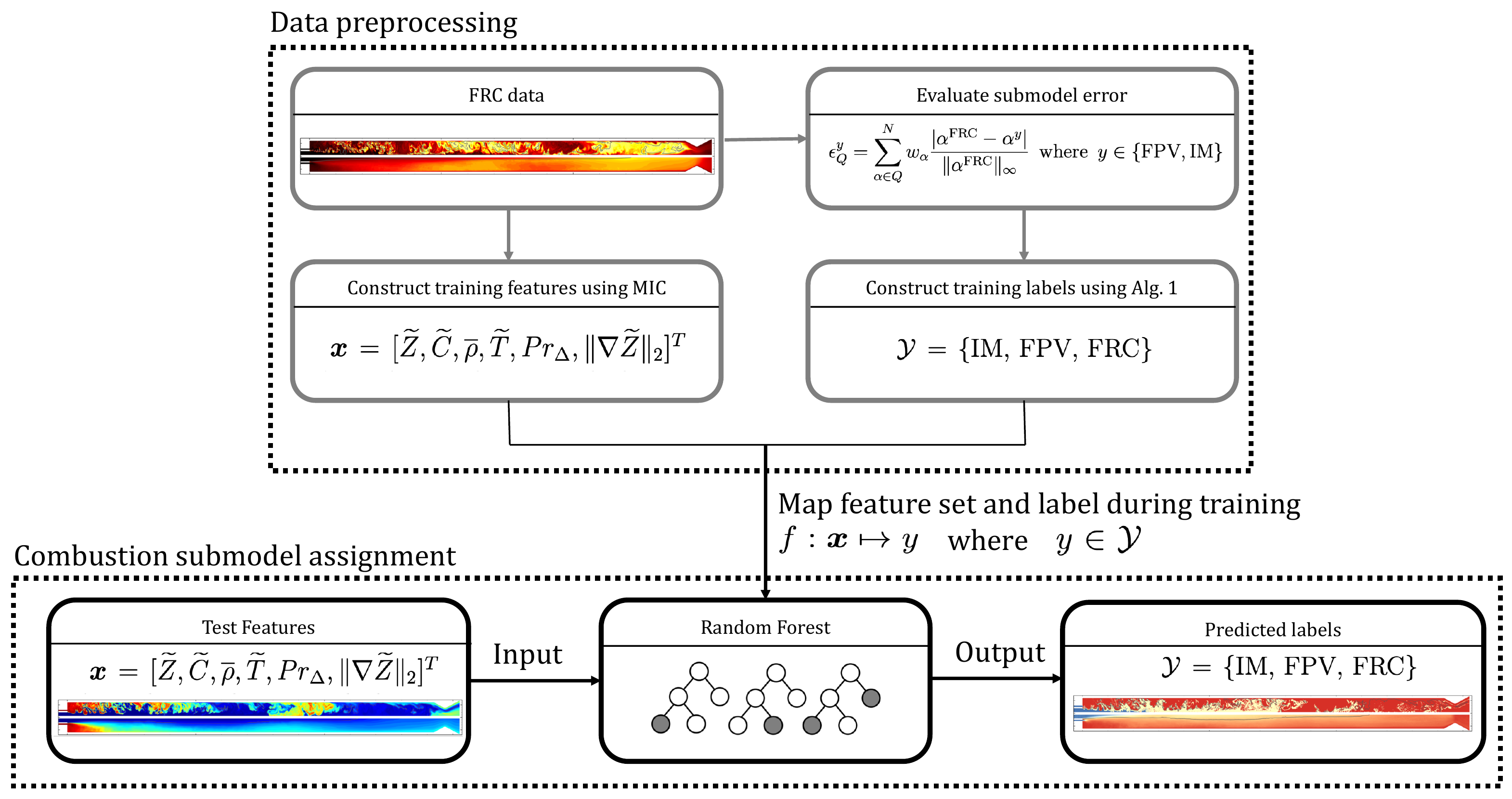}
 \caption{Application of random forest classifier for combustion submodel assignment of a single element GOX/GCH4 rocket combustor. }\label{fig:machLearn}
\end{figure}

\Cref{sec:framework_featurelabel,sec:feature} detailed the procedures applied in this study for preprocessing the monolithic FRC LES data for training. During training, the classification algorithm learns a function $f: \xvec \mapsto y$ that associates  the input vector $\xvec \in \mathcal{X}$, with the corresponding response $y \in \mathcal{Y}$. After training, the learning algorithm can be used to predict the optimal combustion submodel when given new sets of input vectors  $\xvec \in \mathcal{X}$. These steps are summarized in  \cref{fig:machLearn}. 

In this study, we employ the random forest as our classification algorithm.
Random forests~\cite{Breiman2001RandomForests} consist of an ensemble of decorrelated Classification And Regression Trees (CARTs)~\cite{Breiman1984ClassificationTrees}. CARTs are a machine learning approach for formulating prediction models from data by recursively partitioning the inputted feature space, and fitting a simple prediction within each final partition. As a result, the partitioning can be represented graphically as a decision tree. Such decision trees are a graph algorithm, where each node represents a selected feature or attribute, each edge represents a decision based on the properties of this feature, and the leaf nodes represent a final outcome or classification. Decision trees are non-parametric and can model arbitrarily complex relations without any \emph{a priori} assumptions.

In a machine learning algorithm, the expected generalization error is a key characteristic, measuring the accuracy in making predictions for previously unseen data. This error can be decomposed into bias, variance and noise. 
The bias in the predictions is the deviation from the true value of the expectation  (or mean) of the model predictions. In this context, the variance is the variability in the predictions of models. Noise is the inherent stochastic noise in the data.
Decision trees are prone to overfitting. In terms of the bias-variance decomposition, these overfitted models possess low bias but high variance. Ensemble methods offer a simple amelioration by introducing random perturbations in the training procedure to produce several randomized models from the same data, and then combining the predictions of the individual models to form the ensemble prediction. The decorrelated nature of each constituent model reduces the variance of predictions while retaining the low bias.

Random forests are an ensemble method, using ensembles of trees to create a forest. Here, the ensemble model is a collection of  Classification And Regression Trees. The final prediction of this ensemble model is via a majority vote of trained individual trees. The key motivation is to create an ensemble model that has lower variance than the individual trees, while maintaining the low bias. It can be shown that the variance of the ensemble model is directly proportional to the correlation between individual models in the ensemble~\cite{Breiman2001RandomForests}. Thus, the more uncorrelated our individual models are, the lower the variance of the ensemble model. To inject this decorrelation between the individual decision trees in the Random Forest, two concepts are utilized, explicitly:
 \begin{itemize}
     \item Bagging~\cite{Breiman2001RandomForests}: Bagging (or Bootstrap aggregating) is an approach to create different machine learning models from the same data set. In the first step, we can generate multiple new training datasets from the original by sampling from it, uniformly and with replacement (Bootstrapping). Each of these sampled datasets can be used to train a machine learning model. The final prediction is chosen by aggregating the predictions of these individual models (aggregating). 
     In Random forests, each individual tree gets such a bootstrap sample of the original training dataset to learn from. This ensures that every tree has to train on a different dataset and, thus imparts a level of decorrelation to the individual trained tree based models in the ensemble.
     
     \item Random subsampling over features \cite{amit1997joint}: During their training, CARTs are grown by learning splits (or partitions) at each node. Herein, the trees have to determine the best split over the entire set of features to partition the solution space. In random forests, only a small randomized subset of the total set of features is assigned to each tree during training. This introduces additional decorrelation between the trees in the ensemble.
 \end{itemize}
 
 Using Bagging in conjunction with random subsampling over the features, introduces adequate decorrelation over the individual trees in the ensemble to reduce the variance, while maintaining the low bias. In prior investigations, it has been observed that random forests outperform many other algorithms in classification over scalar inputs from structured datasets~\cite{fernandez2014we,wyner2017explaining}.

In the present investigation, the random forest classifier from the {\sc{OpenCV}} library~\cite{Bradski2000TheLibrary} is used. Classification cost scales with  the number of trees, tree depth and the number of training points~\cite{Breiman1984ClassificationTrees}. 
Hence, a random forest consisting of twenty decision trees, and maximum depth of ten nodes is employed. Additionally, $1 \times 10^4$ training points have been randomly sampled from a single LES snapshot consisting of $2\times10^5$ cells. A similar approach is used in other supervised learning problems \cite{Wu2018Physics-informedFramework}.
We must note that the flow in the present configuration is statistically stationary, and thus training data from a single snapshot was found to be sufficient for representing the thermophysical behavior of the combustor.
The number of trees, tree depth, and the number of training points are determined \emph{a priori} by ensuring that the classification performance remains unchanged on a validation set. Training is performed once \emph{a priori}, and requires 530 ms of walltime with 1 CPU. In \emph{a posteriori} simulations, random forest evaluations for $2\times10^5$ cells at each timestep require 1 ms of wall time with 600 CPUs. 
\section{Results} \label{sec:results}

This section assesses the random forest classifier as a method for combustion submodel assignment in data-assisted simulations. \emph{A priori} assessment is performed first to investigate the behavior of random forests when targeting different QoIs. This is followed by an \emph{a posteriori} assessment to study improvements in target QoIs and other quantities that result from the use of random forests in transient data-assisted simulations. \Cref{tab:cases} summarizes  the eight cases, with different QoIs and combustion submodel error threshold values $\theta_Q$, explored in both \emph{a priori} and \emph{a posteriori} assessment.  

\begin{table}[htb!]
 \centering
 \footnotesize
 \caption{Cases investigated in the present study.}
 \label{tab:cases}
 \resizebox{\columnwidth}{!}{
 \begin{tabular}{|l| c c c c c c c c|}
 \hline
 Case & $\theta_T$=0.05 &$\theta_T$=0.02  & $\theta_\text{CO}$=0.05 &$\theta_\text{CO}$=0.02 &$\theta_{\{T,\text{CO}\}}$=0.05 & $\theta_{\{T,\text{CO}\}}$=0.02 
 &$\theta_{\{T,\text{CO},\text{OH}\}}$=0.05 & $\theta_{\{T,\text{CO},\text{OH}\}}$=0.02\\ 
  \hline  \hline
 QoI, $Q$ & $\widetilde{T}$  & $\widetilde{T}$& $\widetilde{Y}_{\ce{CO}}$ & 
                               $\widetilde{Y}_{\ce{CO}}$ & $\{\widetilde{T},\widetilde{Y}_{\ce{CO}}\}$ & $\{\widetilde{T},\widetilde{Y}_{\ce{CO}}\}$
                               & $\{\widetilde{T},\widetilde{Y}_{\ce{CO}},\widetilde{Y}_{\ce{OH}}\}$
                            & $\{\widetilde{T},\widetilde{Y}_{\ce{CO}},\widetilde{Y}_{\ce{OH}}\}$\\ 

 Model threshold, $\theta_Q$ & 0.05 &0.02  & 0.05 &0.02 &0.05 & 0.02 &0.05 & 0.02  \\
 Assessment & \emph{A priori} & \emph{A priori} & \emph{A priori} &  \emph{A priori} & \begin{tabular}{@{}c@{}}\emph{A priori,}  \\ \emph{A posteriori}\end{tabular} &\begin{tabular}{@{}c@{}}\emph{A priori,} \\ \emph{A posteriori}\end{tabular} &\emph{A priori} &\emph{A priori}  \\
 \hline
 \end{tabular}
 }
\end{table}

\subsection{\emph{A priori} assessment} \label{sec:apriori}
\begin{figure}[htb!]
        \centering
            \begin{subfigure}{0.425\columnwidth}
        \centering
            \includegraphics[width=\columnwidth]{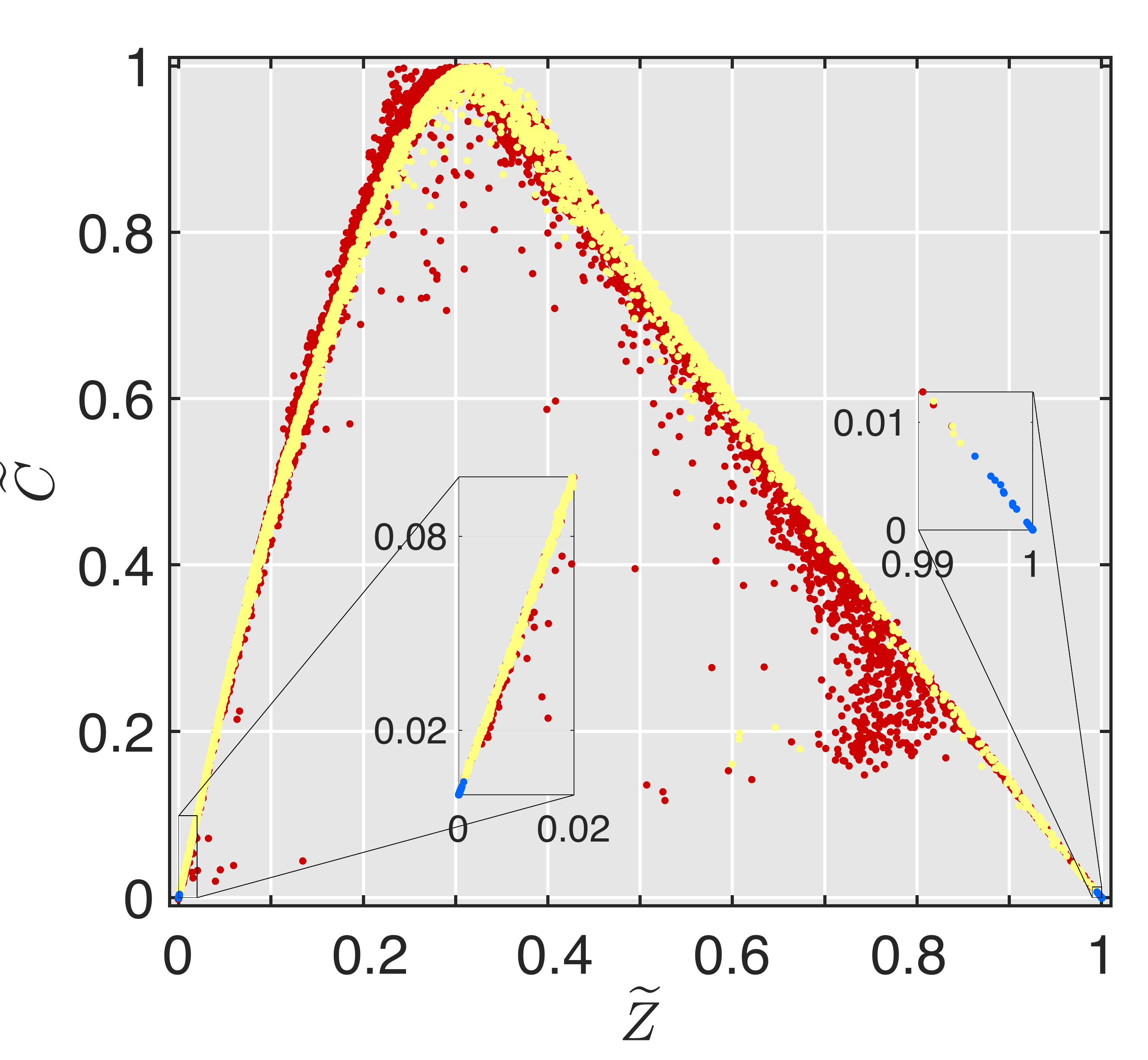}
        \caption{ $\theta_{\{T,\text{CO}\}} = 0.02$}
    \end{subfigure}
    \begin{subfigure}{0.425\columnwidth}
        \centering
            \includegraphics[width=\columnwidth]{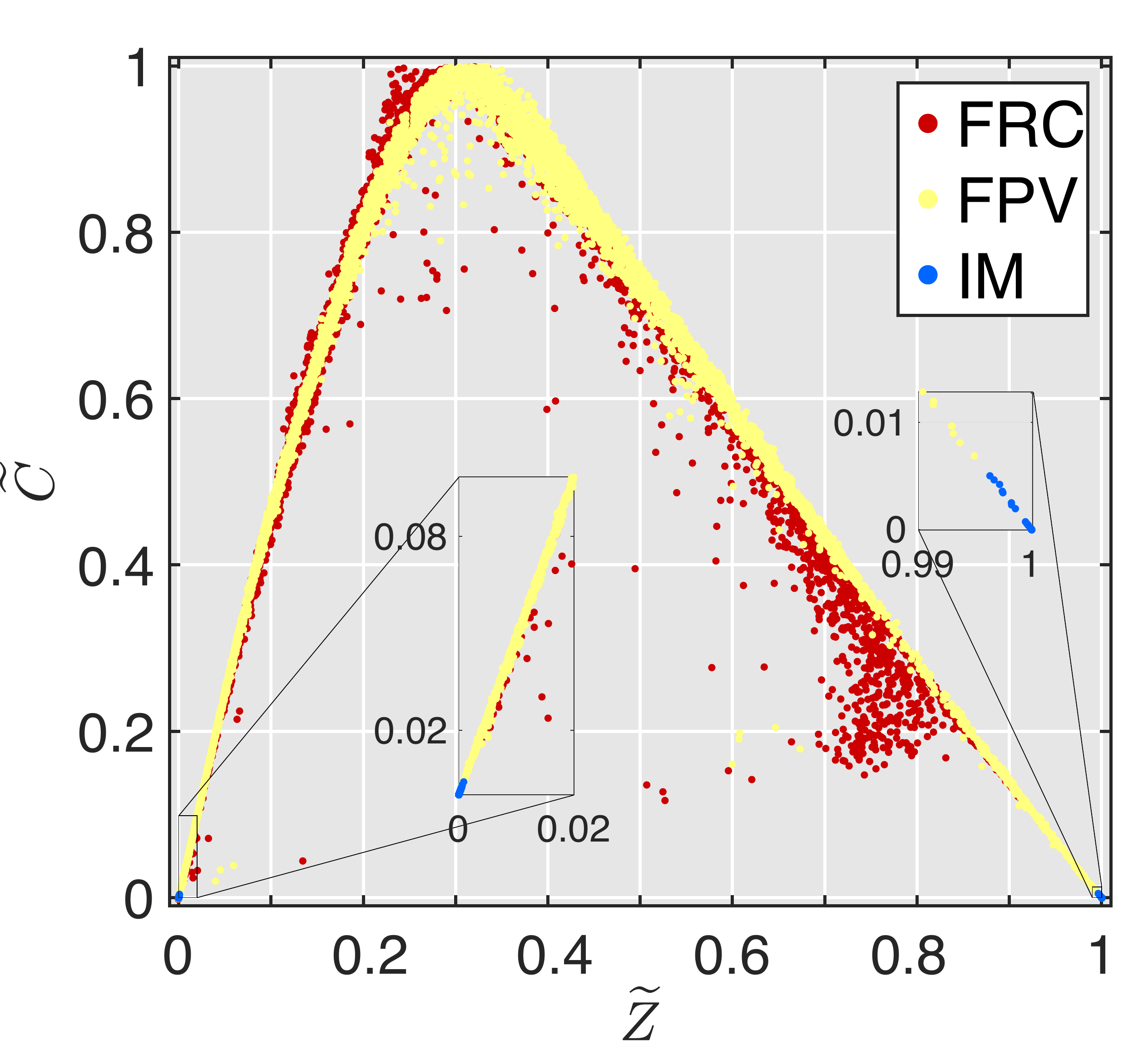}
            \caption{ $\theta_{\{T,\text{CO}\}} = 0.05$}
    \end{subfigure}
    \caption{ Training data for two different combustion submodel error thresholds $\theta_{\{T,\text{CO}\}}$\label{fig:train}.}
\end{figure}

\emph{A priori} assessment involves using the random forest classifier to assign suitable combustion submodels in a test dataset that is created from a monolithic FRC simulation at an unseen timestep. Temperature and  CO and OH mass fraction $\alpha \in \{\widetilde{T},\widetilde{Y}_{\ce{CO}},\widetilde{Y}_{\ce{OH}}\}$ in the test set is then used as QoI for reconstructing the true response, through the procedure described in~\cref{sec:framework_featurelabel}, for comparison with random forest predictions. 
\Cref{fig:train} shows the use of this labeling approach on the training data in $\widetilde{Z}$-$\widetilde{C}$ composition space for $\theta_{\{T,\text{CO}\}} = 0.02$ and  $\theta_{\{T,\text{CO}\}} = 0.05$, respectively. In both cases, IM is shown to be assigned at points where $\widetilde{C}\approx0$, FPV is assigned mostly to conditions near the equilibrium composition. The submodel assignment reverts back to FRC in regions dominated by non-equilibrium effects and heat-losses that are not captured by the adiabatic steady-state flamelet formulation. Employing $\theta_{\{T,\text{CO}\}} = 0.02$  is seen to be more stringent than employing $\theta_{\{T,\text{CO}\}} = 0.05$, with a 0.18 greater fraction of scatter data on the stable branch assigned as FRC, especially for fuel-rich mixtures.
It should be noted that while most out-of-flamelet regions would be assigned FRC, some regions with low reactivity and far from stoichiometry (eg. $\widetilde{Z} = 0.7$)  generate smaller errors which are then be assigned FPV.

\begin{figure}[!htb!]
  \centering
  \includegraphics[width = \columnwidth]{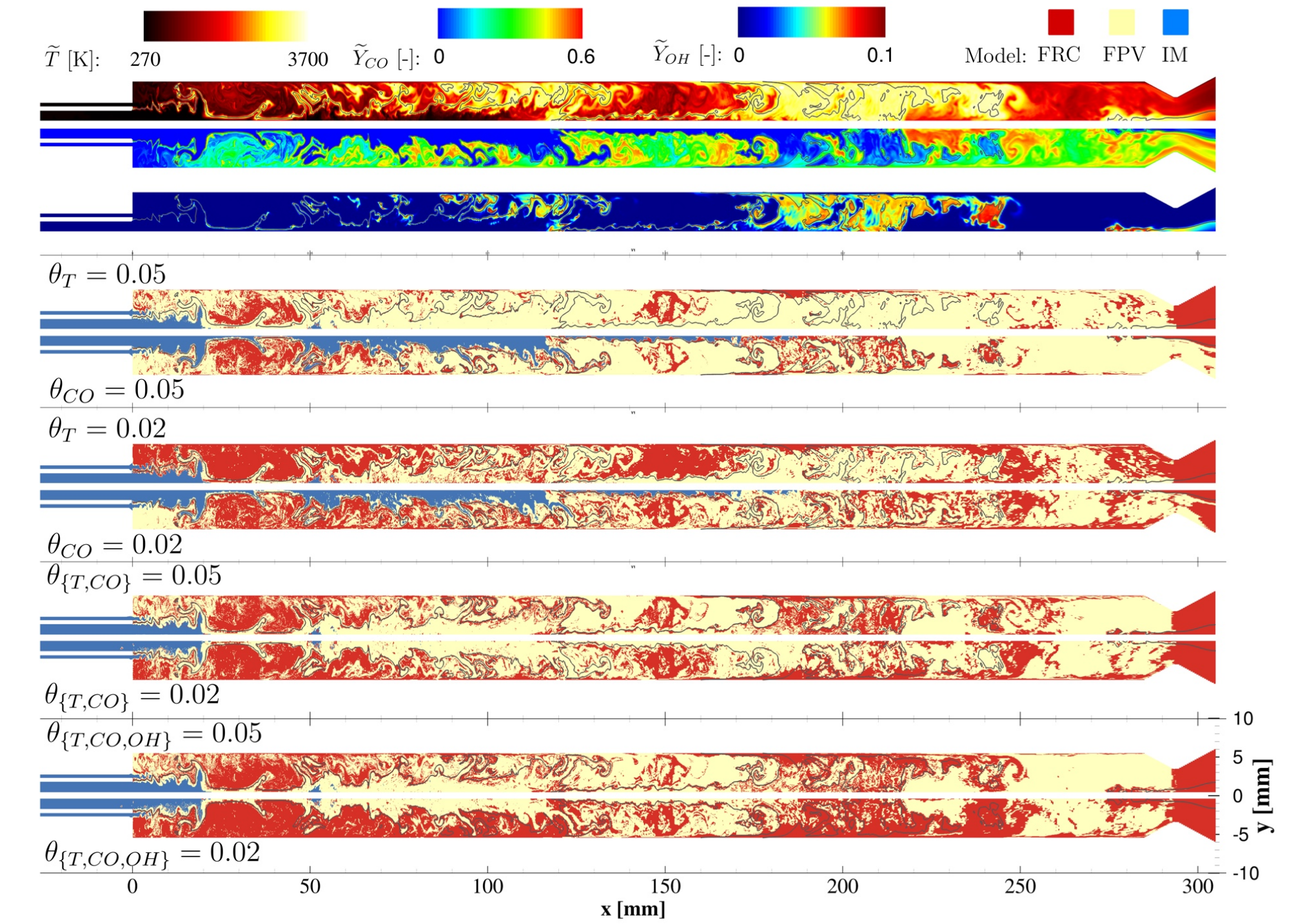}
  \caption{\emph{A priori} analysis, comparing combustion model assignments. Instantaneous temperature, and mass fractions of CO and OH of the test set are also presented; stoichiometric isocontour with $\widetilde{Z}_{st}=0.2$ is shown in black.}\label{fig:priori}
\end{figure}

\Cref{fig:priori} demonstrates the \textit{a priori} combustion submodel assignment on an unseen FRC-simulation snapshot using the six different random forest cases summarized in \cref{tab:cases}. For all six cases, IM is assigned at the injector and the  oxidizer core. In general, FRC is assigned at the near-wall and fuel-rich regions within the combustor where intermediate reactions are not captured well by tabulated chemistry submodels. Using temperature as QoI and a model threshold of $\theta_T = 0.05$ results in an IM assignment of 5\% of the domain, 28\% FRC assignment, with the rest being described by the FPV model.  Constraining the  temperature model threshold $\theta_T = 0.02$ results in FRC assignment in 62\% of the domain, with IM assignment remaining unchanged. 

Using $\widetilde{Y}_{\ce{CO}}$ as QoI and a model threshold of $\theta_{\text{CO}} = 0.05$ results in greater (18\% of the domain) IM assignment, since the CO mass fraction in most of the oxidizer core is close to zero. FRC is assigned to 34\% of the domain. Reducing the CO model threshold $\theta_{\text{CO}} = 0.02$ results in 47\% FRC assignment, with IM assignment unchanged.  Finally the combined use of both temperature and CO mass fraction as QoI,  $Q=\{\widetilde{T},\widetilde{Y}_\text{CO}\}$, results in submodel assignment with combined characteristics of employing each individual QoI.  $\theta_{\{T,\text{CO}\}} = 0.05$ results in 31\% FRC assignment within the domain, while $\theta_{\{T,\text{CO}\}} = 0.02$ results in 52\% FRC assignment. 
Adding OH mass fraction to the QoI set $Q=\{\widetilde{T},\widetilde{Y}_\text{CO},\widetilde{Y}_\text{OH}\}$ increases the FRC assignment to 37\% and 70\% for thresholds $\theta_{\{T,\text{CO},\text{OH} \}} = 0.05$ and $\theta_{\{T,\text{CO},\text{OH} \}} = 0.02$ respectively.
Results demonstrate that reducing model threshold $\theta_Q$ and increasing the number of QoIs increases submodel assignment of FRC.
The submodel assignments for each case are summarized in~\cref{tab:accuracyApriori}.

\Cref{tab:accuracyApriori} also summarizes the true classification of random forests for the eight different cases. Here, true classification is defined as the percentage of classifier assignments that correctly match the true output responses evaluated directly from simulation data.
The true classification fraction range from approximately 0.7 to 0.8, which is comparable to the use of random forests on another classification problem in a flow physics context \cite{Ling2015EvaluationUncertainty}. 
Higher true classification can be achieved through the use of complex deep learning classifiers, which requires (i) more elaborate efforts than the random forests in hyperparameter tuning and (ii) much larger datasets for good performance, and should be subject to further study.

\begin{table}[htb!]
 \centering
 \caption{\emph{A priori} analysis of classifier, summarizing submodel assignment and assignment accuracy.}
 \label{tab:accuracyApriori}
 \resizebox{\columnwidth}{!}{
 \begin{tabular}{|l| c c c c c c c c|}
 \hline
 Case & $\theta_T$=0.05  &$\theta_T$=0.02 & $\theta_\text{CO}$=0.05 &$\theta_\text{CO}$=0.02 &$\theta_{\{T,\text{CO}\}}$=0.05  & $\theta_{\{T,\text{CO}\}}$=0.02 
 &$\theta_{\{T,\text{CO},\text{OH}\}}$=0.05  & $\theta_{\{T,\text{CO},\text{OH}\}}$=0.02 \\
\hline\hline
IM:FPV:FRC  & 5:67:28 & 5:33:62 & 18:48:34& 18:35:47 & 6:63:31 & 6:42:52 & 6:57:37 &6:24:70\\
 True Classification & 0.774 & 0.725 &0.756 &  0.715 & 0.753 &0.734 &0.709&0.691 \\
 \hline
 \end{tabular}}
\end{table}

From \cref{fig:priori}, we observe that model assignment in all six cases is not spatially smooth,  and that model assignment appears speckled. 
This is because  the smoothness of classification boundaries formed within the 6-dimensional feature space is not translated when transformed to physical space.
This is a common issue in classification problems involving spatial data, such as in medical imaging or image processing. Two strategies can be employed to improve spatial smoothness in classification problems \cite{Schindler2012AnClassification,WU_SEE_WANG_IHME_CF2015}: (i) applying the classification techniques to a neighborhood of cells, or (ii) applying a spatial filter on the predicted labels and discretizing the filtered labels. In the \textit{a posteriori} assessment in \cref{sec:aposteriori}, we apply the latter strategy since it is better suited with the current framework that uses local quantities as QoIs and features.  
 
These results demonstrate that the present data-assisted framework enables a fully adjustable level of simulation fidelity through the use of varying submodel error threshold values. Random forests  are  demonstrated to be a reasonably accurate and simple approach for the combustion submodel assignment problems.

\subsection{\textit{A posteriori} assessment: Data-assisted LES}\label{sec:aposteriori}
Data-assisted (DA) simulations using two different model thresholds, $\theta_{\{T,\text{CO}\}} = 0.05$ and  $\theta_{\{T,\text{CO}\}} = 0.02$ are performed by employing random forest classifiers in-flight during simulation runtime. The discussion from this section also includes comparisons with monolithic FRC and FPV simulations.

\begin{figure}[!htb!]
 \centering
 \begin{subfigure}[t]{\columnwidth}
  \centering
  \includegraphics[width = \columnwidth]{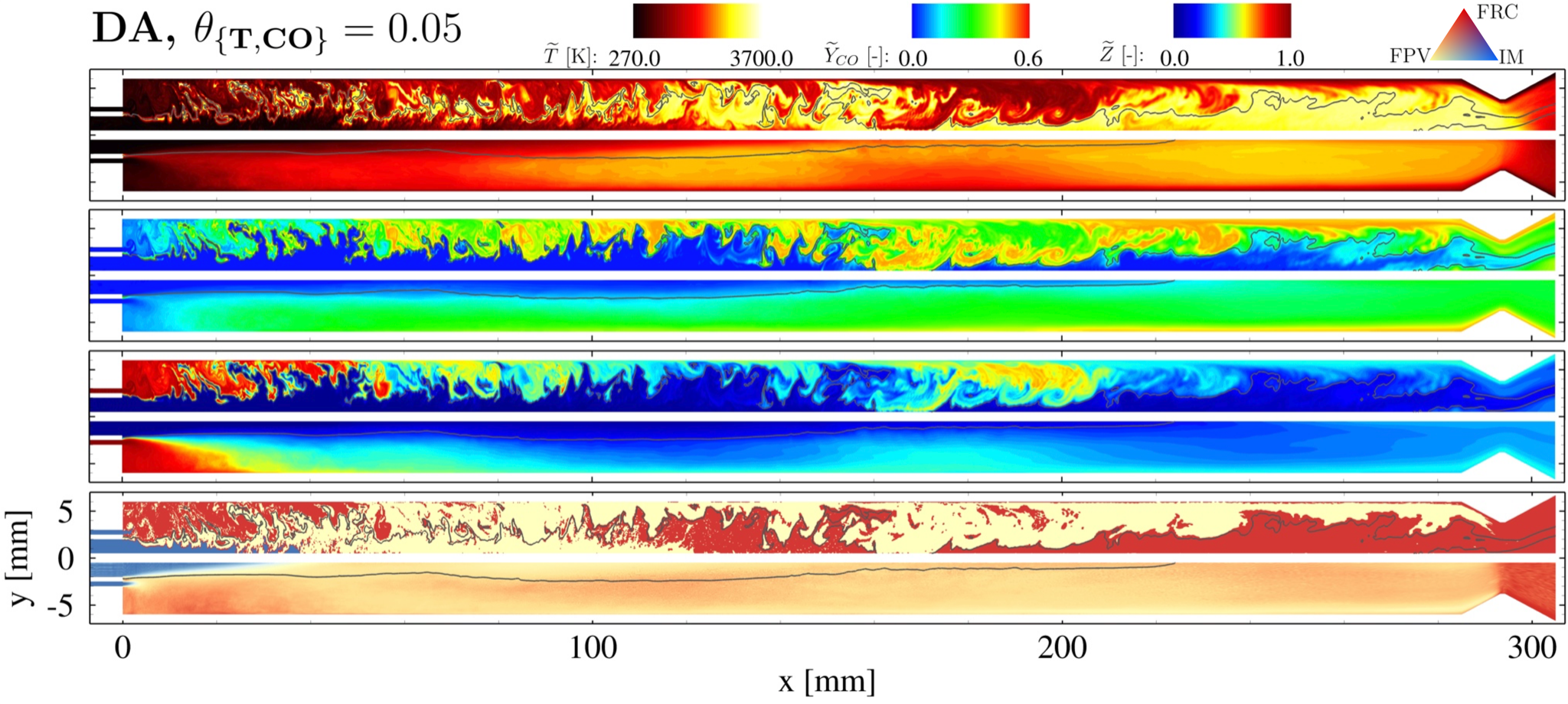}
  \caption{\label{fig:posterioriT5CO5} \emph{A posteriori} DA LES with $\theta_{\{T,\text{CO}\}}=0.05$.}
 \end{subfigure}
 \quad
 \begin{subfigure}[t]{\columnwidth}
  \centering
  \includegraphics[width = \columnwidth]{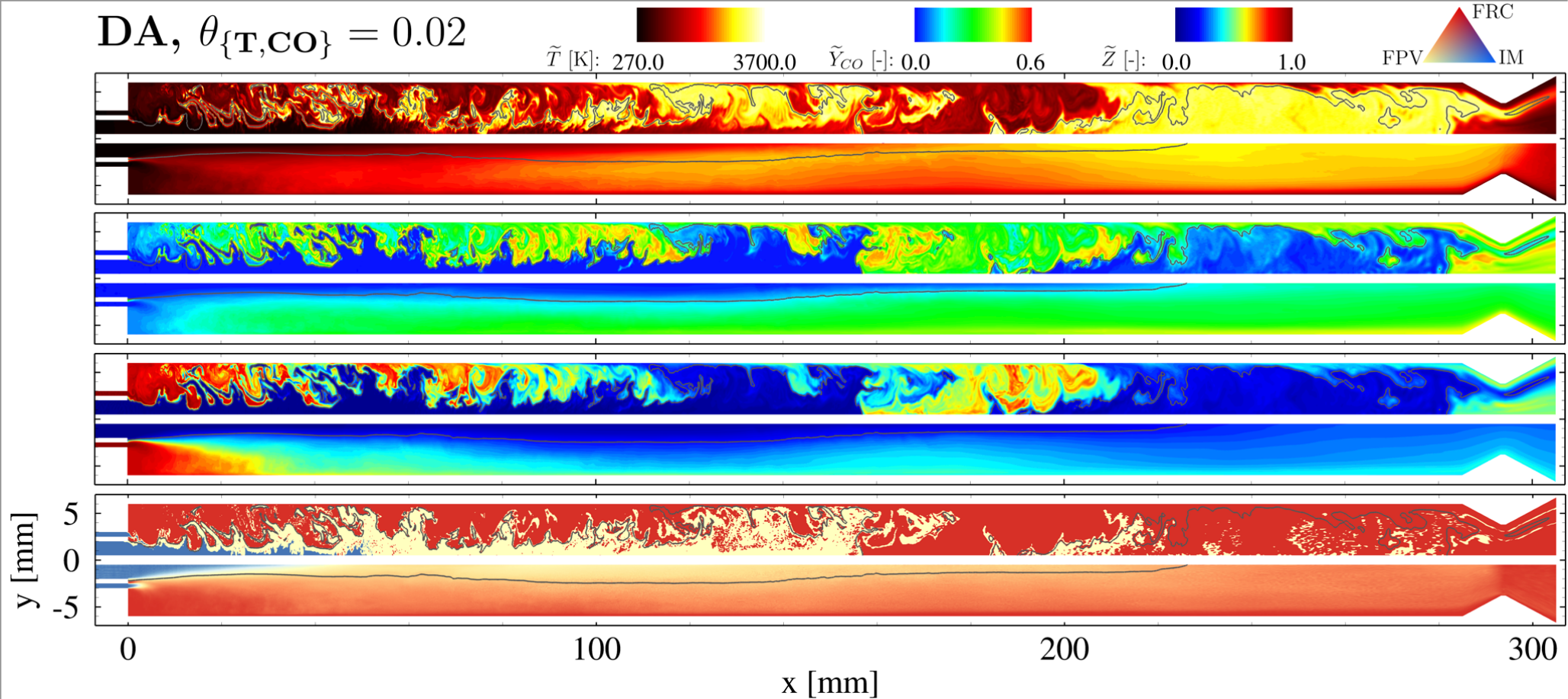}
  \caption{\label{fig:posterioriT2CO2} \emph{A posteriori} DA LES with $\theta_{\{T,\text{CO}\}}=0.02$.}
 \end{subfigure}
 \caption{\label{fig:posterioriDA}Temperature, CO mass fraction, and mixture fraction fields (from top to bottom) from \emph{a posteriori} DA LES for (a) $\theta_{\{T,\text{CO}\}}=0.05$ and (b) $\theta_{\{T,\text{CO}\}}=0.02$. Upper half: instantaneous fields, bottom half: time-averaged fields; stoichiometric isocontour with $\widetilde{Z}_{st}=0.2$ is shown in black.}
\end{figure}

\Cref{fig:posterioriT5CO5} shows that employing  model threshold $\theta_{\{T,\text{CO}\}} = 0.05$ on the DA simulation results in temperature predictions that are in good agreement with the monolithic FRC simulation, shown in~\cref{fig:posterioriFRC}. However, time-averaged results show that a thin layer of CO  develops at the chamber wall at 170 mm. Additionally, a thicker thermal boundary layer is also observed when compared to monolithic FRC simulations. Nonetheless, both species and thermal boundary layers are thinner than the monolithic FPV simulations that were presented in~\cref{fig:posterioriFPV}. Averaged FRC utilization with $\theta_{\{T,\text{CO}\}}=0.05$ is at 34\% of the domain with IM-utilization at 4\%.
In addition, a thin intermittent area close to the wall is also assigned FRC. This indicates that the random forest recognizes the importance of wall effects on CO and temperature  but that the user-defined model error threshold  $\theta_{\{T,\text{CO}\}}=0.05$ is too large.

\Cref{fig:posterioriT2CO2} shows that tightening the model threshold $\theta_{\{T,\text{CO}\}} = 0.02$ results in temperature, CO, and mixture fraction fields that agree with the monolithic FRC simulation, shown in~\cref{fig:posterioriFRC}. Model assignment using this threshold results in 60\% FRC utilization. Before $x=150$ mm FRC is assigned to all fuel-rich and near-wall regions. For $x>150$ mm, FRC is assigned to most of the domain where incomplete combustion products and intermediate species are dominant.

\begin{figure}[htb!]
 \centering
 \begin{subfigure}{0.425\columnwidth}
  \includegraphics[width=\textwidth]{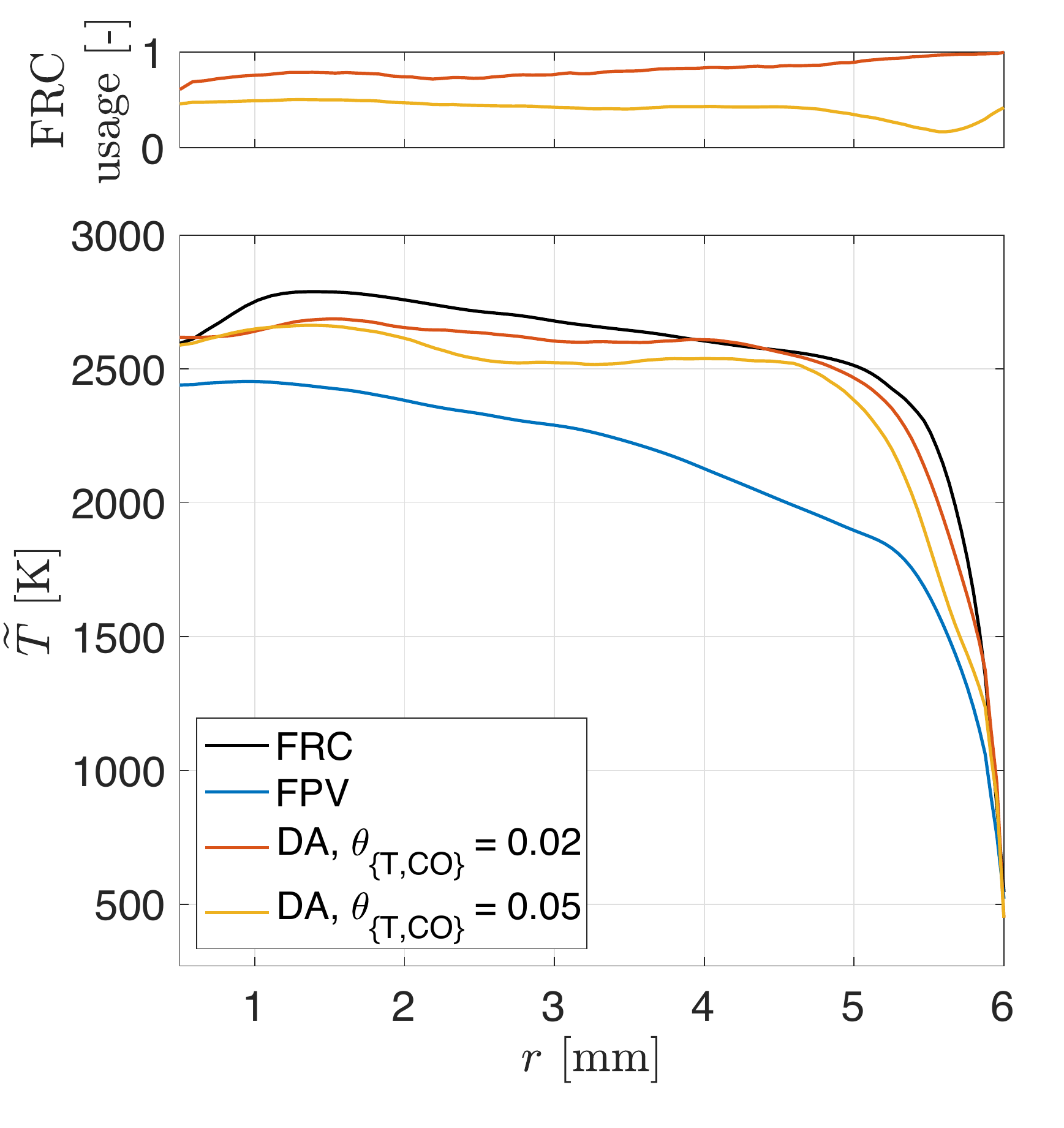}
  \caption{Temperature}
 \end{subfigure}
 \begin{subfigure}{0.425\columnwidth}
  \centering
  \includegraphics[width=\textwidth]{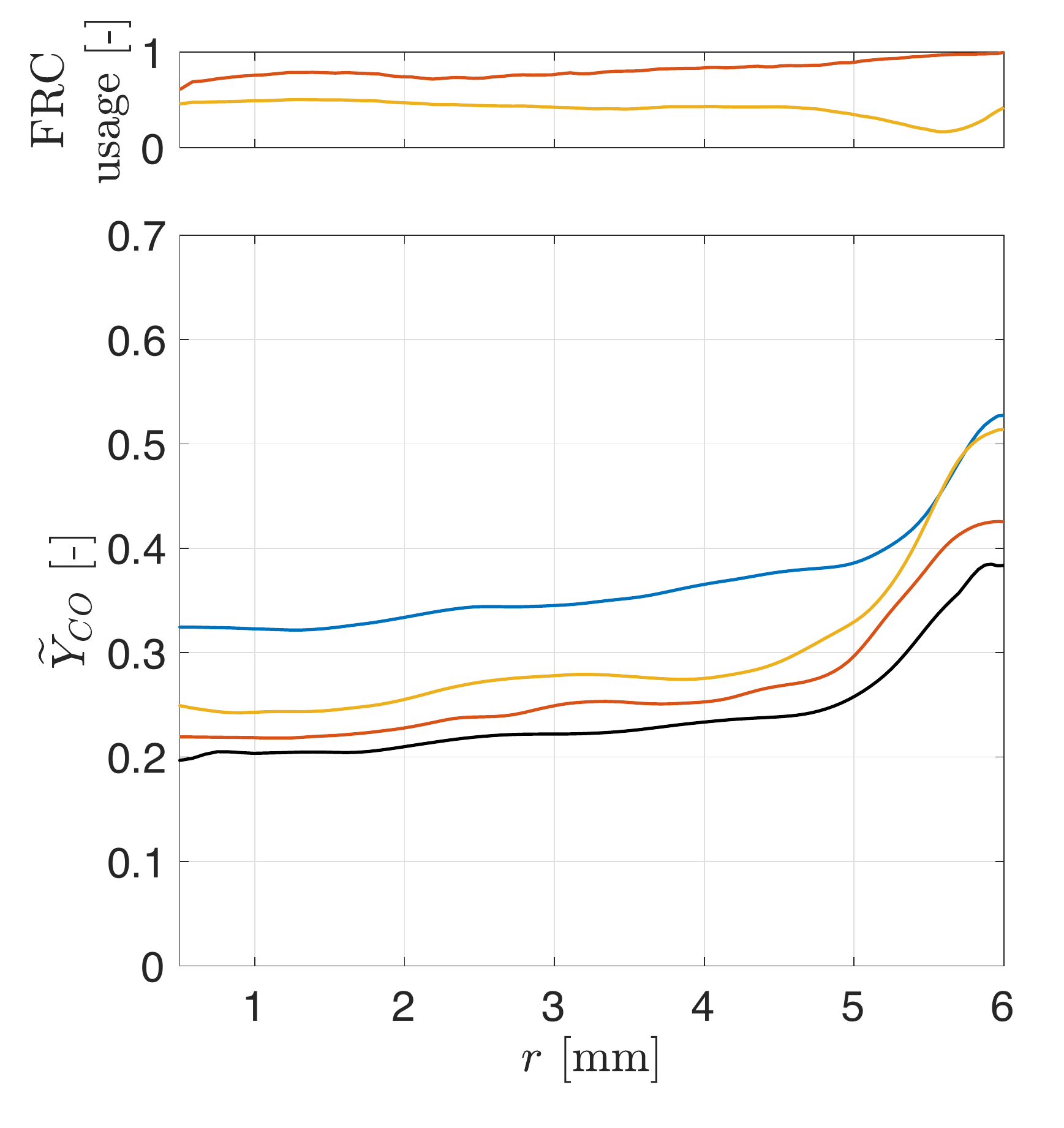}
  \caption{CO mass fraction}
 \end{subfigure}
 \caption{Comparisons of time-averaged radial profiles of (a) temperature and (b) CO mass fraction between monolithic FRC, monolithic FPV, and data-assisted (DA) simulations at an axial distance $x=250$ mm. Time-averaged utilization of FRC is included. \label{fig:radialPlot}}
\end{figure}

\Cref{fig:radialPlot} shows comparisons of radial profiles of time-averaged temperature and CO mass fraction at an axial distance of 250 mm.  Effects of wall-heat loss on the monolithic FPV simulation  is seen to reduce the overall temperature and thicken the thermal boundary layer, which in turn results in greater CO mass fraction. Using a model threshold of $\theta_{\{T,\text{CO}\}}=0.05$, DA-predictions for temperature and CO mass fraction profiles away from the wall are in good agreement with monolithic FRC simulations, and averaged FRC submodel utilization ranges between 16\% and 38\%. 
At $r=5$ mm, the random forest is able to recognize when the absolute error between temperature diminishes and thus assigns less FRC accordingly, which results in greater temperature and CO mass fraction deviation from monolithic FRC simulations. After $r=5.7$ mm, the random forest begins to recognize the importance of near-wall effects and assigns more FRC. However, this FRC utilization is still insufficient for recreating monolithic FRC simulations. Further constraining the DA-simulation threshold to $\theta_{\{T,\text{CO}\}}=0.02$ improves the agreement with monolithic FRC-simulations. However, small errors can still be seen even with high FRC submodel utilization that ranges from 61\% to 90\%. 

\begin{figure}[!htb!]
        \centering
            \includegraphics[width=97mm]{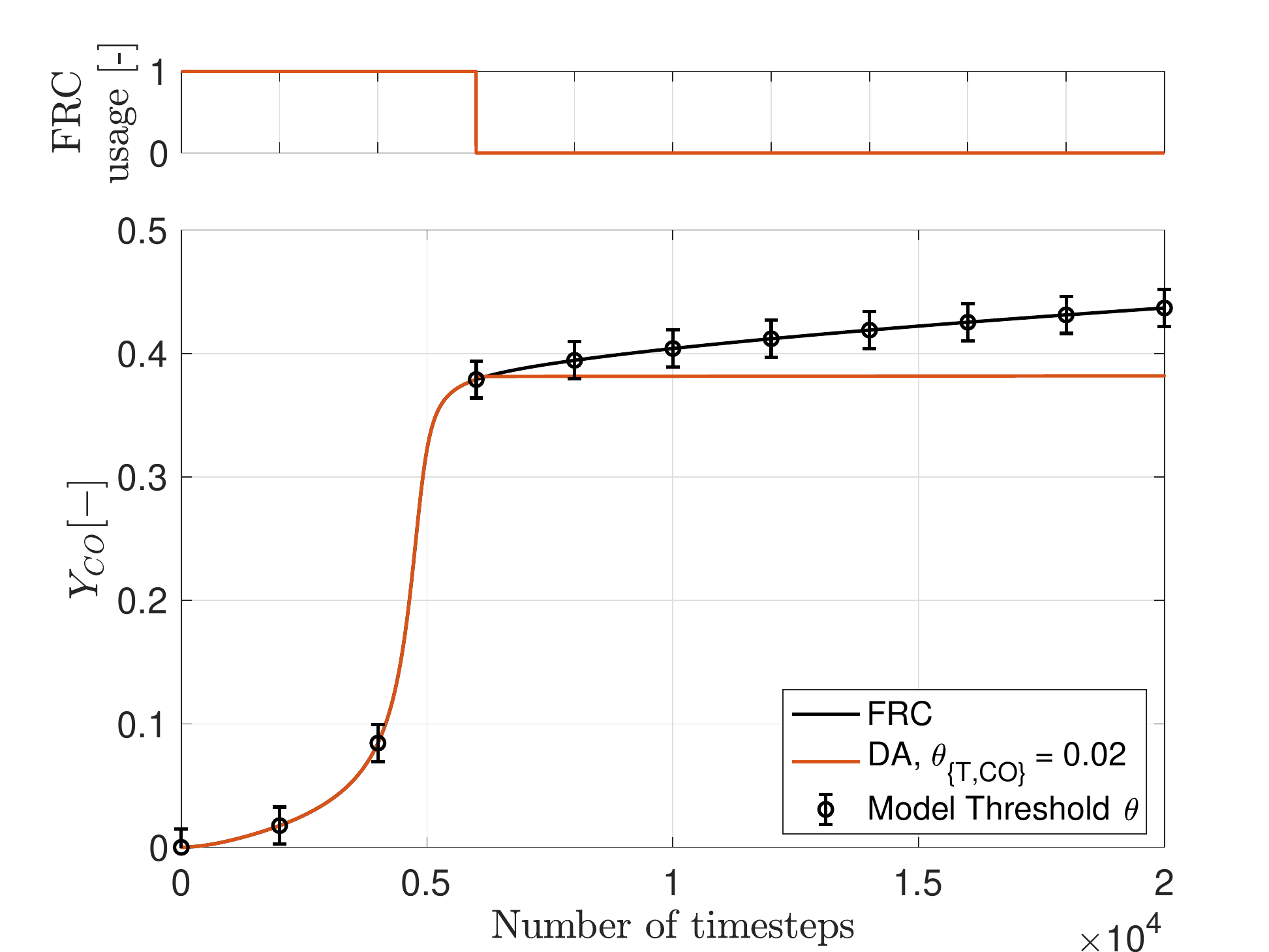}
     \caption{FRC and DA-assisted calculation of CO mass fraction as a function of time step in a 0D homogeneous reactor. }\label{fig:0d}
\end{figure}

Results from \cref{fig:radialPlot} show that the present data-assisted modeling approach can generate simulation results that are in agreement with monolithic FRC calculations. However errors observed are greater than the local model error threshold $\theta_{\{T,\ce{CO}\}}$ used for training the random forests. This is caused by small changes in one state that can result in significant deviations in  later states. 
This effect is illustrated by applying DA combustion modeling with local model error threshold $\theta_{\{T,\text{CO}\}}=0.02$ on CO mass fraction, using a rich methane-air mixture ($Z=0.55$) in a constant pressure homogeneous reactor at 20 bar and initial temperature of 1800~K, as shown in  \cref{fig:0d}.  
In this setup, it is observed that while the random forest correctly assigns the correct model based on local model error at 5800 timesteps, the CO trajectory leads to a total error exceeding the local error threshold of 0.02 as the DA simulation no longer has knowledge of the monolithic FRC CO production beyond this timestep and cannot recover to the correct state.
However, the benefit of the present approach is that, in  the worst-case, errors made do not exceed errors made by the lowest fidelity combustion model employed.

\begin{figure}[htb!]
\centering
  \begin{subfigure}{0.425\columnwidth}
   \includegraphics[width=\textwidth]{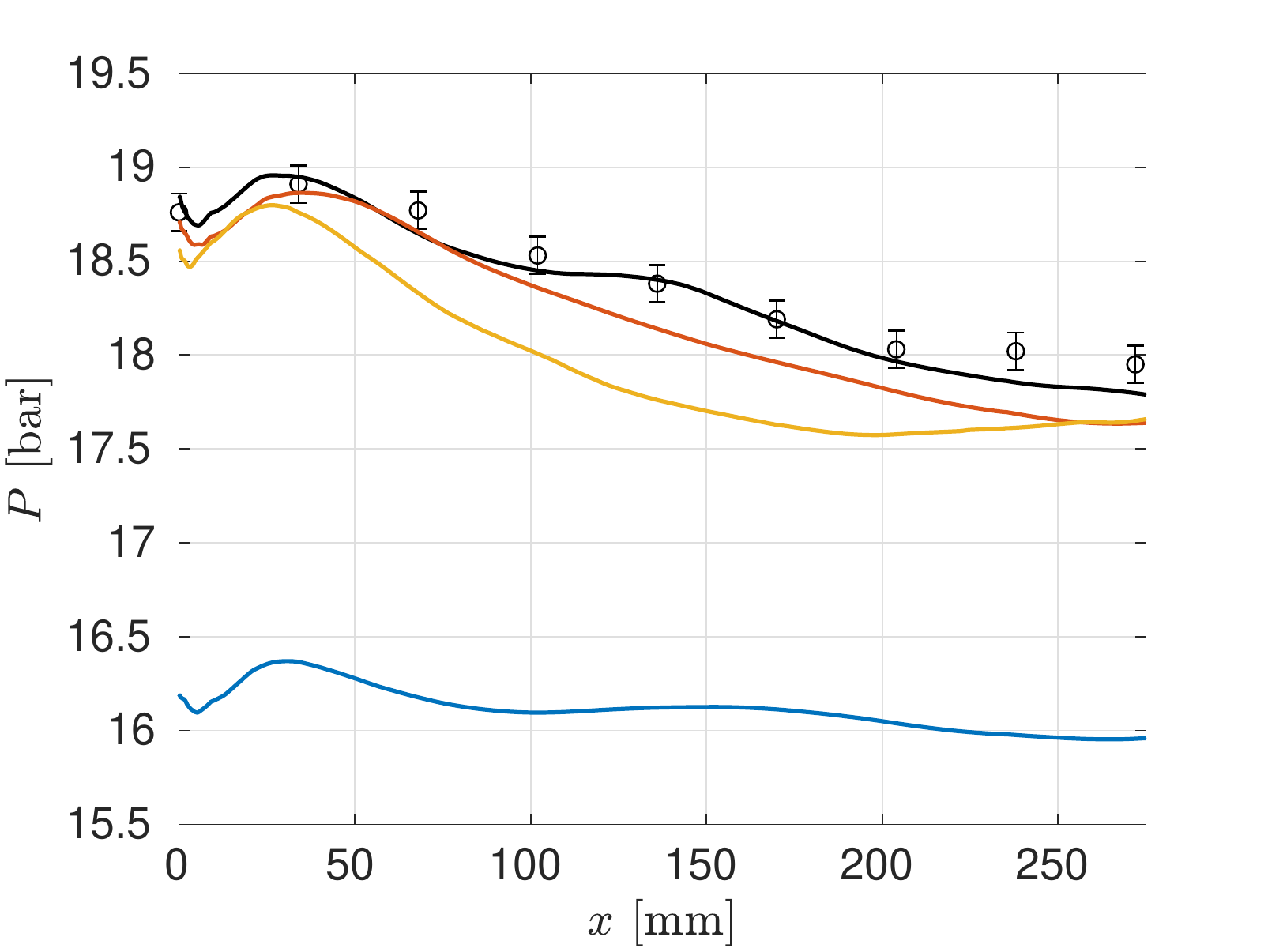}
   \caption{\label{fig:exp_p}Wall pressure}
  \end{subfigure}
  \begin{subfigure}{0.425\columnwidth}
   \centering
   \includegraphics[width=\textwidth]{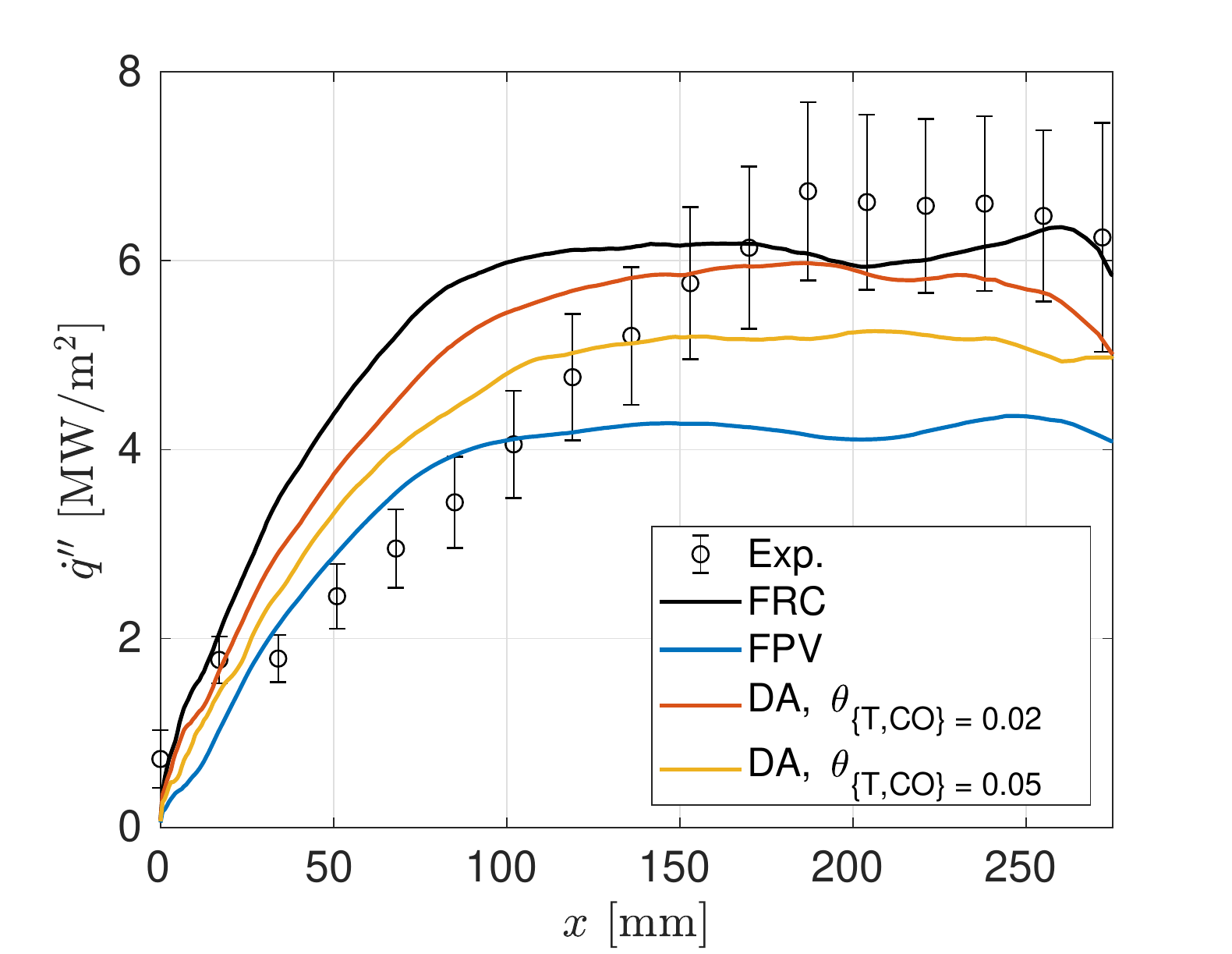}
   \caption{\label{fig:exp_qw}Wall heat flux}
  \end{subfigure}
  \caption{ Comparison of simulation results for (a) wall pressure and (b) wall heat flux calculations with experimental measurements~\cite{Perakis2019InverseChambers}. \label{fig:exp}}
\end{figure}

Generating numerical predictions that match experimental wall measurements are challenging for this rocket combustor case, since these quantities are dependent on overall flow and temperature fields in a highly nonlinear system. Studies \cite{Muller2016NumericalStudy,PERAKIS_HAIDN_IHME_PCI2021} comparing LES and RANS results have reported up to 8\% deviation from wall pressure measurements. Wall heat flux predictions are more sensitive to simulation parameters, where deviations up to 75\% have been reported in the same studies. While the aim of the present study is not to find simulation results that match the experimental results, LES calculations of wall pressure and wall heat flux are presented with measurements by~\citet{Perakis2019InverseChambers} in~\cref{fig:exp} to quantify effects of applying the DA formulation on overall combustor behavior.

\Cref{fig:exp_p} shows that wall pressure predictions between monolithic FRC agree well with experimental measurements. The DA simulation with $\theta_{\{T,\text{CO}\}}=0.02$  shows a small underprediction, but still possesses reasonable agreement with monolithic FRC. The DA simulation with $\theta_{\{T,\text{CO}\}}=0.05$ shows a greater underprediction. Wall pressure underprediction can be caused by reduced fuel conversion~\cite{Roth2016NumericalCombustor}. This is likely the case since higher CO levels in both cases are observed in~\cref{fig:radialPlot}.  Additionally, the monolithic FPV simulation also demonstrates the lowest pressure and highest CO levels.

\Cref{fig:exp_qw} shows that wall heat flux predictions for FRC simulation are in good agreement with experimental data after $x=120$ mm, but with a steeper heat flux rise. This steep heat flux rise is likely due to the misrepresentation of turbulent mixing in a thin axisymmetric domain, and is also seen in other axisymmetric studies \cite{Lapenna2018SimulationMethod,Zips2017Non-AdiabaticCombustion}. Tightening the model threshold $\theta_{\{T,\text{CO}\}}$ results in better convergence with monoolithic FRC calculations.
The DA simulation with $\theta_{\{T,\text{CO}\}}=0.02$  is in reasonable agreement with the FRC simulation, while the FPV simulation demonstrates the lowest heat flux due to low overall temperatures from low combustion efficiency.

\begin{figure}[!htb!]
  \centering
  \includegraphics[width=0.625\columnwidth]{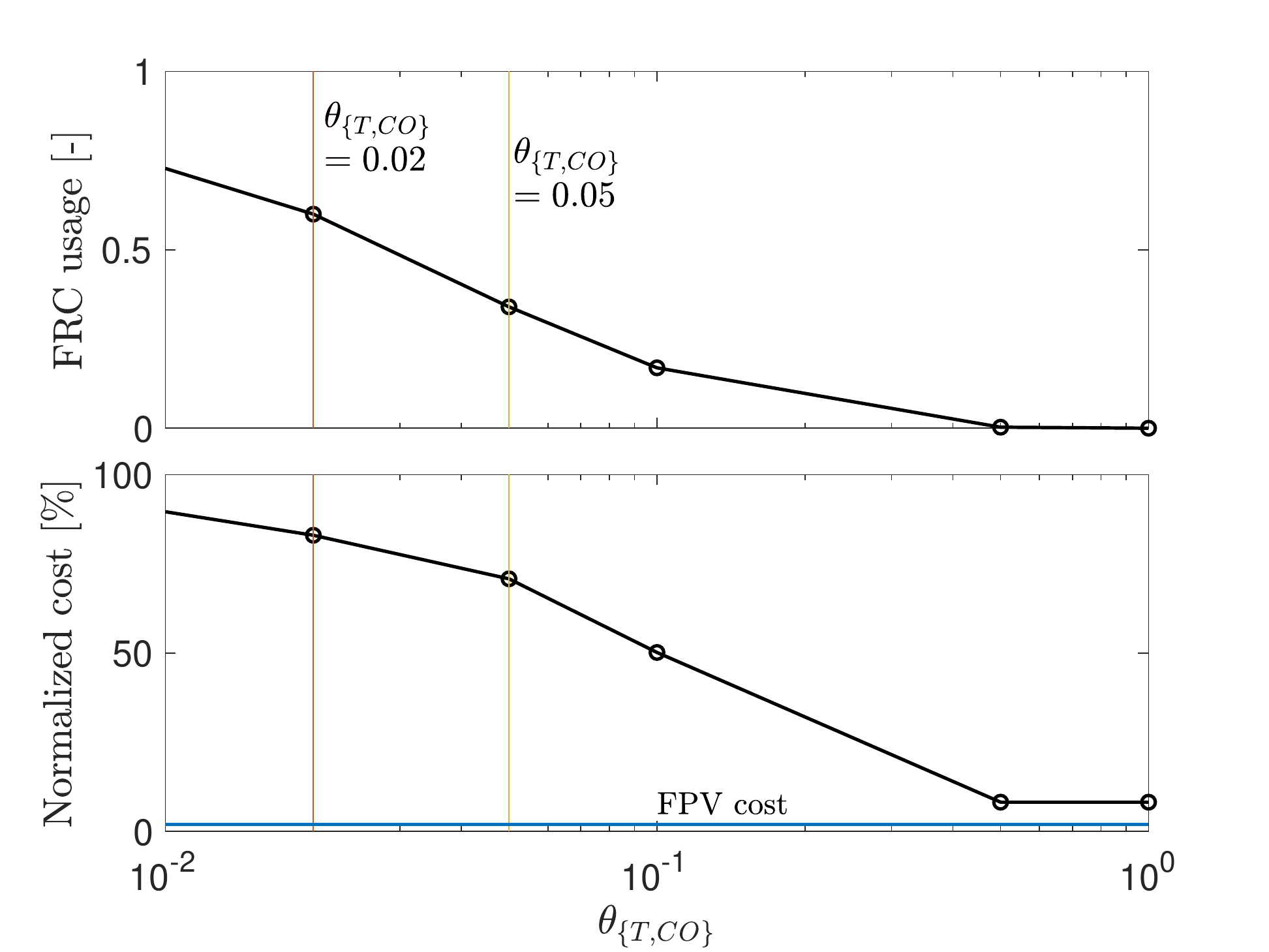}
  \caption{FRC utilization and normalized computational cost versus combustion submodel error threshold $\theta_{\{T,\text{CO}\}}$. }\label{fig:cost} 
\end{figure}

\Cref{fig:cost} shows FRC usage and corresponding computational cost (normalized by FRC cost) of the data-assisted simulation as a function of combustion submodel error threshold $\theta_{\{T,\text{CO}\}}$ when computed using 600 Intel Xeon (E5-2680v2) processors. Each timestep in the FPV simulation requires 50 ms of wall time to solve, while each timestep in the FRC requires a wall time of 2,300 ms. When $\theta_{\{T,\text{CO}\}}=0.50$, the classifier does not assign FRC in the entire domain,  resulting in a normalized cost of 8\%. This additional cost represents the overhead from the random forest evaluation and the coupling of the three combustion submodels in the same domain.
Simulations performed in this study utilized 34\% ($\theta_{\{T,\text{CO}\}} = 0.05$) and 60\% FRC ($\theta_{\{T,\text{CO}\}} = 0.02$), which resulted in 70\% and 80\% of FRC cost, respectively. 
These results demonstrate that classification algorithms can be utilized in high-fidelity simulations to reduce computational cost. Further reductions of the computational cost is achievable by combining the method proposed in this work with regression techniques~\cite{CHATZOPOULOS20131465,Franke2017} to reduce the complexity of the finite-rate chemistry representation.

\subsection{Generalization}
\begin{figure}[!htb!]
 \centering
  \includegraphics[width = \columnwidth]{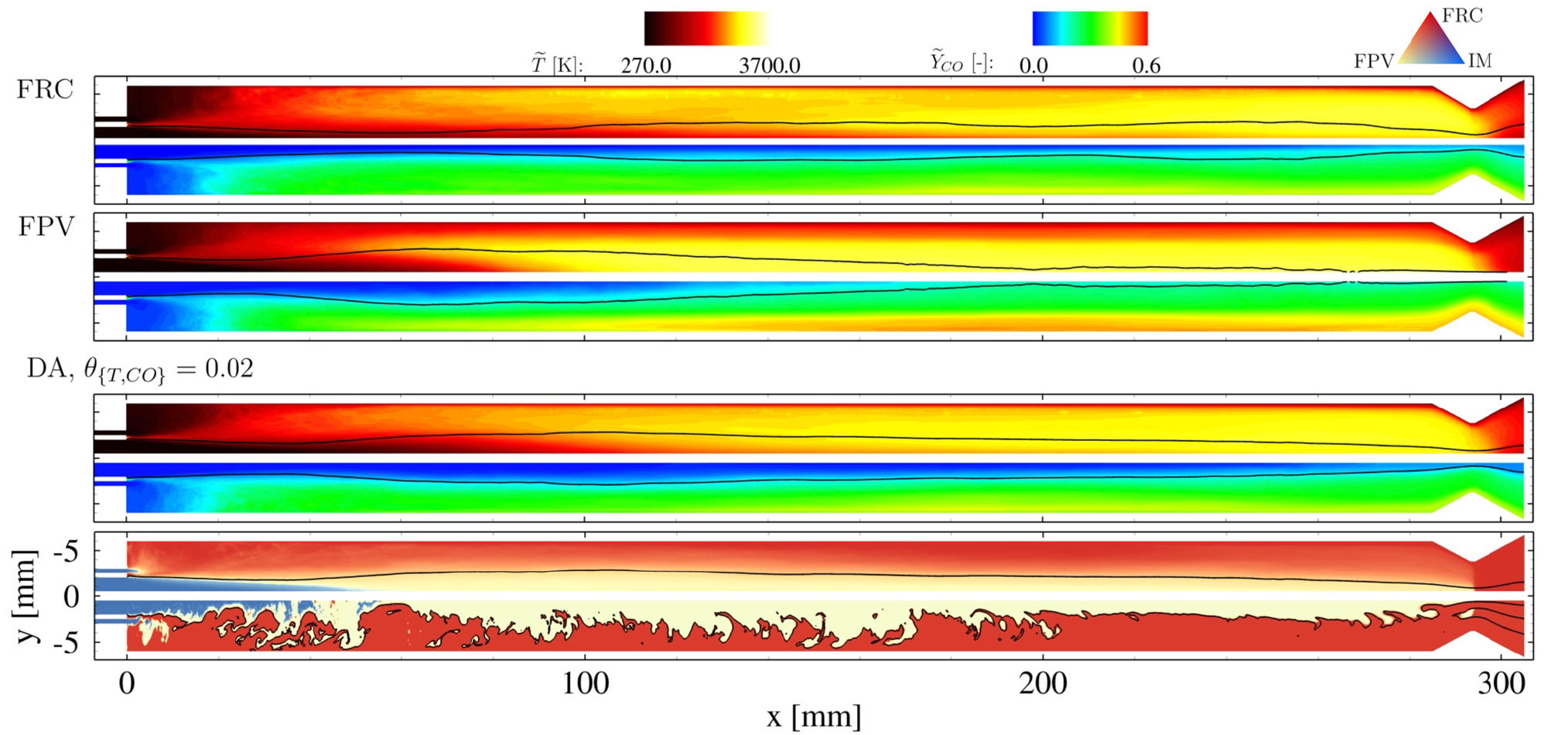}
 \caption{\label{fig:posteriori3x}Comparison of time-averaged temperature and CO mass fraction fields for monolithic FRC, monolithic FPV, and  \emph{a posteriori} DA LES ($\theta_{\{T,\text{CO}\}}=0.02$) on a configuration with three times the inlet mass flow rate. Time-averaged and instantaneous model assignment for DA LES is shown at the bottom. Stoichiometric isocontour with $\widetilde{Z}_{st}=0.2$ is shown in black.}
\end{figure}

In order to demonstrate the ability of random forests to generalize, additional LES are performed on a modified configuration with three times the inlet mass flow, while keeping all other parameters constant.
\Cref{fig:posteriori3x} compares time-averaged temperature and CO mass fraction fields for monolithic FRC, monolithic FPV, and  \emph{a posteriori} DA LES ($\theta_{\{T,\text{CO}\}}=0.02$) for this setup.
All three LES cases in this modified configuration demonstrate a longer oxygen core than the original configuration (\cref{fig:posterioriFRC_FPV}) due to higher flow velocity, indicating less complete combustion. When compared to FRC, FPV overpredicts the thickness of the thermal boundary layer and CO formation. 
DA LES with model threshold  ($\theta_{\{T,\text{CO}\}}=0.02$) predicts temperature and CO flow fields in good agreement with monolithic FRC calculations. 
 Random forest assigns FPV to the lean side of the flame, while assigning FRC to the rich side. This is also seen in the DA case of the original configuration in \cref{fig:posterioriDA} from 0 to 150 mm, where major combustion products have not fully formed.
  Model assignment using this threshold results in 51\% FRC and 6\% IM utilization, resulting in 77\% of the FRC cost. 
  
  Results from this modified configuration demonstrate that the present data-assisted approach can be applied to different configurations as long as the training data can represent the underlying thermo-physical behavior. We note that all simulations and training data from the present study employ the same mesh. Since the random forest classifies well in this modified configuration,  this method should still be effective for different  mesh resolutions as long as the flow can be represented by local points of the training data. The generalizability of this method improves with increasing availability of representative data.

\section{Conclusions} \label{sec:conclusions}
This study introduced a data-assisted modeling approach, employing random forest classifiers, as a method for dynamic and local combustion model assignment in reacting flow simulations.
\textit{A priori} assessment was conducted on the random forests, which were fed with six input features based on local thermofluid properties, to evaluate the behavior of the classifiers during submodel assignment when targeting different QoIs. 
Random forests were shown to assign three different candidate combustion models -- finite-rate chemistry (FRC), flamelet progress variable (FPV) approach, and inert mixing (IM) -- based on predefined QoIs with fraction of true classification ranging from approximately 0.70 to 0.80. 

Two cases of \textit{a posteriori} simulations using random forest classifiers for combustion submodel assignment  during simulation runtime, were performed. 
Time-averaged results  of temperature and \ce{CO} mass fraction demonstrated that the data-assisted simulation produced  species and temperature profiles in better agreement with monolithic FRC than monolithic FPV calculations. The use of the random forest with submodel error threshold of $\theta_{\{T,\text{CO}\}}=0.02$ results in significant improvements from monolithic FPV simulations in all quantities at a 20\% lower cost than monolithic FRC calculations. An additional DA LES ($\theta_{\{T,\text{CO}\}}=0.02$), performed on a modified configuration with three times the inlet mass flow rate, demonstrated that the present approach can be applied to different configurations as long as the training data can represent the relevant thermo-physical behavior.

Results from \textit{a priori} and \textit{a posteriori} assessments demonstrated that the present data-assisted framework is adjustable and effective for the purpose of combustion model assignment, so long as high-quality data is available.
While this method avoids the challenging task of constructing a mathematical model-compliance indicator \cite{WU_SEE_WANG_IHME_CF2015}, the present approach is not Pareto-optimized since only local submodel errors were utilized for training.
Thus, additional concepts from the Pareto-efficient combustion framework can supplement the present data-assisted LES framework.
Additionally, the exploration of other cost-efficient and accurate classification algorithms could improve the classification accuracy of the present data-assisted approach. 
In particular, ANNs with deep learning architectures have shown high accuracy in numerous classification problems.
Other opportunities for extending this work include 
 (i) the extension of the current framework to bridge local submodel error with  non-local errors, (ii) the addition of non-local quantities in the feature and label set, and (iii) the consideration of a more extensive candidate combustion submodel set. 

\section*{Acknowledgments}
The authors gratefully acknowledge financial support from the Air Force Office of Scientific Research under Award No. FA9300-19-P-1502, NASA with award No. 80NSSC18C0207, and Stanford University Harold and Marcia Wagner Engineering Fellowship.
Resources supporting this work are provided by the High-End Computing (HEC) Program at NASA Ames Research Center.
\bibliography{paperDACombustion.V01.bib} 
\bibliographystyle{elsarticle-num-PROCI_titlemod.bst}


\end{document}